\documentclass[alpha-refs]{wiley-article}

\usepackage{graphicx}
\usepackage[space]{grffile}
\usepackage{subcaption}
\usepackage{latexsym}
\usepackage{textcomp}
\usepackage{longtable}
\usepackage{tabulary}
\usepackage{booktabs,array,multirow}
\usepackage{amsfonts,amsmath,amssymb}
\usepackage{natbib}
\usepackage{url}
\usepackage{hyperref}
\usepackage[capitalise,noabbrev]{cleveref}
\usepackage{xcolor}
\hypersetup{colorlinks=true,linkcolor=blue,citecolor=blue,urlcolor=blue}
\usepackage{etoolbox}
\usepackage[normalem]{ulem}

\newtheorem{definition}[theorem]{Definition}
\newtheorem{remark}[theorem]{Remark}
\newtheorem{assumption}[theorem]{Assumption}

\AtBeginDocument{\DeclareGraphicsExtensions{.pdf,.PDF,.eps,.EPS,.png,.PNG,.tif,.TIF,.jpg,.JPG,.jpeg,.JPEG}}
\usepackage[utf8]{inputenc}
\usepackage[english]{babel}

\usepackage{listings}
\definecolor{codebg}{RGB}{248,248,248}
\definecolor{codeframe}{RGB}{200,200,200}
\definecolor{keyword}{RGB}{0,0,180}
\definecolor{comment}{RGB}{0,128,0}
\definecolor{string}{RGB}{163,21,21}
\lstdefinelanguage{SysMLv2}{
  keywords={package, part, port, item, attribute, connection, flow, state,
            action, requirement, constraint, verification, analysis,
            view, viewpoint, concern, enum, abstract, private, public,
            import, def, exhibit, perform, first, then, transition,
            accept, connect, to, from, in, out, doc, variation, variant,
            allocation},
  keywordstyle=\color{keyword}\bfseries,
  sensitive=true,
  morecomment=[l]{//},
  morecomment=[s]{/*}{*/},
  morestring=[b]",
  commentstyle=\color{comment}\itshape,
  stringstyle=\color{string},
}
\crefname{lstlisting}{Listing}{Listings}
\Crefname{lstlisting}{Listing}{Listings}
\crefname{theorem}{Theorem}{Theorems}
\crefname{proposition}{Proposition}{Propositions}
\crefname{lemma}{Lemma}{Lemmas}
\crefname{definition}{Definition}{Definitions}
\crefname{remark}{Remark}{Remarks}
\crefname{assumption}{Assumption}{Assumptions}
\Crefname{assumption}{Assumption}{Assumptions}

\papertype{Original Article}
\paperfield{Systems Engineering}
\corraddress{Sai Sandeep Damera, Department of Electrical and Computer Engineering, University of Maryland, College Park, MD 20742, USA}
\corremail{sdamera@umd.edu}
\fundinginfo{}

\title{A SysML~v2 Domain Library for ROS~2 Robot Architectures}
\author[1]{Sai Sandeep Damera}
\author[1]{John S.\ Baras}
\affil[1]{Department of Electrical and Computer Engineering, Institute for Systems Research, University of Maryland, College Park, MD 20742, USA}
\runningauthor{Damera and Baras}

\makeatletter
\zref@newlabel{mdf@pagelabel-1}{\mdf@pagevalue{1}}
\makeatother
\begin{document}
\maketitle

\begin{abstract}
Systems engineers who model a ROS2 robot in SysML~v2 must first model ROS2 itself. Its nodes, communication endpoints, message types, lifecycle states, quality-of-service policies, and coordinate frames have no ready definitions in the language, so each project reconstructs them. We present \texttt{ros2-sysmlv2}, a native SysML~v2 domain library that supplies these definitions for reuse. A robot model imports the library and specializes its definitions to declare its own nodes, ports, connections, frames, and parameters. The library maps ROS2 concepts to SysML~v2 definitions and usages and organizes them in Primitives, Infrastructure, and Application layers. We check its declarations for parity with the ROS2~Jazzy sources they model. Because every ROS2 element of a model specializes a library definition, a tool can select those elements by type. We state what this selection is and use it for a round trip. A ROS2 package is generated from the selection, and a monitor generated with the package compares the running graph with the same selection and reports every departure against the model. The generation-gap workflow derives topic endpoints and parameter declarations while preserving developer-owned node algorithms. In a case study on a generated ground robot, the generated conformance monitor reported each of the three departures that we introduced in the implementation. The library is distributed as a versioned package that any SysML~v2 model can import.
\end{abstract}


\section{Introduction}
\label{rsl:sec:introduction}

Robot Operating System~2 (ROS2) systems are assembled from software nodes, communication endpoints, message types, configuration values, and coordinate frames. An engineer who wants this architecture in a system model has to represent each of these concepts first. ROS2 describes them across interface definitions, launch files, parameter YAML, and source code, and it offers no model-level definition of a node, a publisher, or a quality-of-service (QoS) policy that a system model could reuse. Each modeling effort therefore builds its own ROS2 vocabulary. A domain library that defines the vocabulary once, in the modeling language itself, lets every robot model start from shared and checked definitions.

Model-Based Systems Engineering (MBSE) provides a setting for such a vocabulary. SysML is widely used to represent system structure, behavior, requirements, and verification information in a common model \citep{friedenthal2014practical}. SysML~v2 supplies a normative textual notation alongside its graphical notation, a definition and usage style for reusable types and their occurrences, and first-class ports, connections, flows, and state machines \citep{omg2025sysml,omg2025kerml}. These features make a textual SysML~v2 package suitable for a domain library that can be versioned, imported, specialized, and processed by tools.

This paper asks two questions. How can a native SysML~v2 library provide a reusable and semantically disciplined vocabulary for modeling ROS2 systems, and what does a model built on that library give the engineer beyond the vocabulary? We answer the first with \texttt{ros2-sysmlv2},\footnote{Sources at \url{https://github.com/SEIL-ISR/ros2-sysmlv2}, package index page at \url{https://sysand.com/projects/seil-isr/ros2-sysmlv2/}.} a published domain library organized around the corresponding concepts in ROS2. The library uses SysML~v2 constructs directly. ROS2 message types become \texttt{item def}s, node roles become \texttt{part def}s, communication endpoints become \texttt{port def}s, bindings between endpoints become \texttt{interface def}s, and transforms between frames become \texttt{connection def}s. Specialization carries recurring patterns from abstract node archetypes to application-specific instances.

\begin{figure}[!t]
\centering
\includegraphics{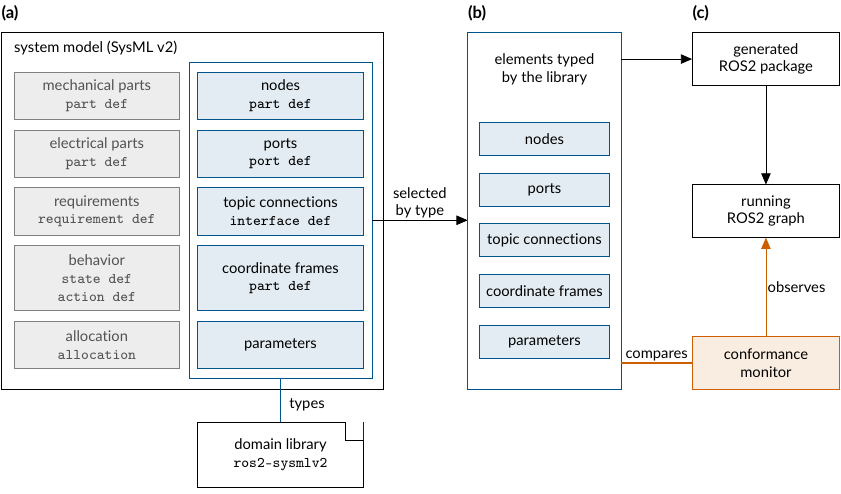}
\caption{A system model built on the library, and the tools that read it. (a) The SysML~v2 model of a robot holds mechanical and electrical parts, requirements, behavior, and allocation (grey boxes) together with nodes, ports, topic connections, coordinate frames, and parameters typed by \texttt{ros2-sysmlv2} (blue boxes). (b) The elements whose types specialize library definitions form the ROS2 projection of the model. (c) A ROS2 package is generated from the projection. The conformance monitor generated with it observes the running graph and compares it with the projection. The grey content stays in the model for its own consumers.}
\label{rsl:fig:overview}
\end{figure}

The library is a vocabulary, and the robot models are written by its users. A robot model imports the library and specializes its definitions. It remains an ordinary SysML~v2 model that can hold whatever else its authors model, such as mechanical parts, requirements, behavior, and allocation. Because every ROS2 element of such a model specializes a library definition, a tool can select those elements by type, and that selection is the whole of what the software side of the model needs. A ROS2 package is generated from it, with its topic endpoints, parameters, and lifecycle in place and its node algorithms left to the developer. A monitor generated with the package compares the running ROS2 graph with the same selection and reports each departure against the model. We answer the second question with this round trip, shown in \cref{rsl:fig:overview}, and with two results that state what the selection is and when a running graph conforms to it.

We make four contributions:
\begin{enumerate}
    \item A native SysML~v2 mapping for ROS2 messages, endpoints, node roles, lifecycle state, QoS profiles, coordinate frames, parameters, and node archetypes, with the ownership and referential-feature choices that semantically valid declarations need. The layered, packageable library \texttt{ros2-sysmlv2} carries this mapping and is checked against the ROS2~Jazzy sources it models.
    \item The type-directed projection of a model built on the library, with one stated result. The projection is the largest set of usages with a library type, it is closed under ownership and connection ends and independent of the order in which a tool reads the model, and a reference extractor computes it from the containment tree.
    \item The round trip on the projection. A ROS2 package is generated from it, with node algorithms in developer-owned files, and a monitor generated with the package decides structural conformance of the running graph. One stated result says what conformance is, that every departure names its key and coordinate, and within what bound a persistent departure is reported.
    \item Evidence from use. A ground robot generated from a model built on the library alone, whose monitor reported each of the three departures that we introduced in the implementation, and a downstream library that builds on the published release.
\end{enumerate}

The article has five more parts. \Cref{rsl:sec:related-work} places the work among ROS modeling and MBSE approaches. \Cref{rsl:sec:library} presents the library. \Cref{rsl:sec:projection} defines the software projection, states what it is, and describes the package generated from it. \Cref{rsl:sec:conformance} states when a running graph conforms to the projection and describes the generated monitor. \Cref{rsl:sec:case-study} reports the case study, and \cref{rsl:sec:conclusion} concludes. The supplementary material carries the library in more detail, the proofs, the pipeline and the monitor in detail, and the record of the case study.

\section{Related Work}
\label{rsl:sec:related-work}

MBSE has been applied to robotic systems for trajectory planning, manipulation, simulation-driven grasping, and sensor-suite trade studies \citep{bansal2018model,meehan2021model,sekar2022model,damera2024integrated}, and to the design and integration of complex robotic systems \citep{pasupuleti2023model}. Their models use SysML~v1 tools and graphical workflows. MeROS is a SysML~v1 profile and metamodel for ROS~1 and ROS2 systems \citep{winiarski2023meros}. It represents nodes, topics, services, actions, parameters, and the packages of a developer workspace through stereotypes, with message, service, and action data structures as blocks. ROS2ML is a SysML profile for the communication among nodes and the link and joint structure of a robot \citep{guizani2023new}. \texttt{ros2-sysmlv2} differs from these profiles on three points. It uses the SysML~v2 definition and usage vocabulary itself, so a robot model specializes library definitions and a SysML~v2 validator checks their ownership and typing rules, without a separate profile or metamodel. It models message types field by field, QoS policies and presets, the managed-node lifecycle, and coordinate frames, and scripts compare these declarations with the ROS2 sources. It is distributed as a versioned package that a model imports.

BRIDE demonstrated model-driven ROS~1 development through generated components, node skeletons, and launch files \citep{bubeck2014bride}. SmartMDSD provides a component, port, connector, and deployment environment for service robotics, and ROSMOD generates ROS~1 code from a web-based component model \citep{stampfer2016smartmdsd,kumar2016rosmod}. HAMR generates code for the Java virtual machine, Linux, and the seL4 microkernel from AADL architecture models \citep{hatcliff2021hamr}, and it has since gained a front end for a SysML~v2 subset that represents AADL models \citep{hatcliff2025end}. These systems establish the value of explicit architectural models and generated implementation scaffolding. Their vocabularies are custom metamodels or the concepts of another architecture language, or they target ROS~1. \texttt{ros2-sysmlv2} instead packages a ROS2 vocabulary in the standardized SysML~v2 language and treats code generation as one consumer of the model. HAROS analyzes ROS repositories statically to assess their quality and to study how ROS primitives are used \citep{santos2016framework,santos2017mining}. ROSDiscover recovers the run-time architecture of a ROS system statically and detects misconfigurations in it \citep{timperley2022rosdiscover}, and ROSpec is a language for specifying and verifying the configuration of ROS components \citep{canelas2025rospec}. These analyses and a native architecture library are complementary. An existing deployment can be represented through library instances, while an architecture-first workflow derives implementation scaffolding and a monitor from the same instances.

The SysML~v2 specification and the specification of the Kernel Modeling Language (KerML), on which SysML~v2 is built, define the language foundation and its standard package archive format \citep{omg2025sysml,omg2025kerml}. Syside provides a text-first toolchain and programmatic access to native \texttt{.sysml} sources \citep{vaicenavivcius2025syside}, and Sysand provides a registry and package manager for \texttt{.kpar} archives \citep{sensmetry2026sysand}. The library uses this ecosystem to distribute a reusable vocabulary rather than a one-off project model.

\section{The SysMLv2-ROS2 Library}
\label{rsl:sec:library}

\subsection{A Native Mapping for ROS2 Architecture}
\label{rsl:sec:native-mapping}

A ROS2 system is a set of \emph{nodes}, named units of computation that discover one another at run time and form, with the communication among them, the ROS2 graph. Nodes exchange data through \emph{topics}, named channels that carry messages of one type from publishers to subscribers, through \emph{services}, a request answered by a response, and through \emph{actions}, a goal that takes time, with feedback and a result. Every publisher and subscriber carries QoS policies, and the two are connected only when the publisher offers at least what the subscriber requests. A node is configured through named, typed \emph{parameters}, a managed node follows a standard \emph{lifecycle} whose transitions an external manager requests, and coordinate frames are related by the transforms of the \texttt{tf2} library. This paper uses Jazzy, the long-term-support distribution of 2024, and Nav2, the navigation package of ROS2 \citep{macenski2020marathon}.

The library starts from the architectural objects that engineers name when they design a ROS2 system: message types, endpoints, nodes, their connections, lifecycle state, QoS policies, frames, and parameters. Each object is mapped to a SysML~v2 construct whose semantics match its role. \Cref{rsl:tab:mapping} summarizes the central correspondences. A message type is structured information exchanged at an interface, so it is an \texttt{item def}. A node is a reusable component type, so it is a \texttt{part def}. A publisher or subscriber is an interaction boundary with direction and a message type, so it is a \texttt{port def}. A topic, service, or action binding joins two ports, so it is an \texttt{interface def}, the SysML~v2 construct for a connection between ports. A transform joins two frames, which are parts, so it is a \texttt{connection def}. QoS policies and parameters are configuration values represented by \texttt{attribute def}s.

\begin{table}[htbp]
\centering
\footnotesize
\caption{Native SysML~v2 constructs used for the ROS2 architectural vocabulary.}
\label{rsl:tab:mapping}
\begin{tabular}{@{}p{3.4cm}p{3.4cm}p{6.5cm}@{}}
\toprule
\textbf{ROS2 architectural object} & \textbf{SysML~v2 construct} & \textbf{Library role} \\
\midrule
Message interface & \texttt{item def} & Structured message fields and cardinalities \\
Node role & \texttt{part def} & Node identity, specialization, and owned endpoints \\
Publisher or subscriber & \texttt{port def} & Directed typed communication endpoint \\
Topic, service, or action binding & \texttt{interface def} & Relation between two endpoint roles \\
QoS policy and parameter & \texttt{attribute def} & Configuration value, range, and descriptor \\
Lifecycle behavior & \texttt{state def} & Primary and transition states, requested transitions, and transitions guarded on callback results \\
Coordinate frame & \texttt{part def} and \texttt{connection def} & Frame identity and static or dynamic transform relation \\
\bottomrule
\end{tabular}
\end{table}

The mapping uses the language's definition and usage distinction. Definitions establish reusable types. Usages place those types into a robot architecture and specialize them with concrete names, message types, QoS values, and parameters. The library thereby represents common structure once and permits instances to carry their deployment-specific choices.

This native treatment keeps the vocabulary inside the language. A model that uses the library is an ordinary SysML~v2 model, so its other elements can refer to the ROS2 elements as they refer to any part or port, and the extractor takes the ROS2 elements out of it and leaves the rest in place. The ROS2 constructs are thus declared with the language's own definitions rather than attached through an external profile.

\subsection{Semantic Discipline for Definitions and Usages}
\label{rsl:sec:semantic-discipline}

The choice of a SysML~v2 construct carries ownership and typing consequences. A declaration varies by structural kind, direction, and composition. Kind distinguishes attributes, items, parts, and ports. Direction distinguishes \texttt{in}, \texttt{out}, and \texttt{inout} interaction features and gives a port its endpoint role. Composition distinguishes a value owned by its container from a referential feature whose lifetime is independent of that container. The library applies these dimensions deliberately because a textual spelling that parses can still violate the language's semantic rules.

Message definitions provide the clearest ownership example. A \texttt{PoseStamped} message is composed from a \texttt{Header} and a \texttt{Pose}; its structured fields are therefore composite \texttt{item} usages inside an \texttt{item def}, so the nested data travel with the message. Scalar message data use typed attributes and declared cardinalities. In contrast, a port definition is an interface boundary. Its non-port nested features must be referential under the SysML~v2 port rules. A QoS profile is therefore an \texttt{attribute def} carried by an attribute usage, while a domain value of item type placed inside a referential container uses the \texttt{ref item} form. This preserves the intended domain type while respecting the feature's ownership semantics. Connection ends are referential as well, because a connection relates endpoints that already exist.

The design follows the normative rules for attributes, referential features, items, ports, and connection ends in the SysML~v2 specification \citep{omg2025sysml}: attribute and directed usages are referential, attribute usages are typed by attribute definitions or datatypes, and item usages by item definitions. The library sources and the two robot models of \cref{rsl:sec:case-study} pass the Syside validator \citep{vaicenavivcius2025syside}, version 0.11.0, with warnings treated as errors, so a declaration that parses but breaks one of these rules is reported.

\subsection{Layers and Packages}
\label{rsl:sec:layers}

\begin{figure}[!t]
\centering
\includegraphics{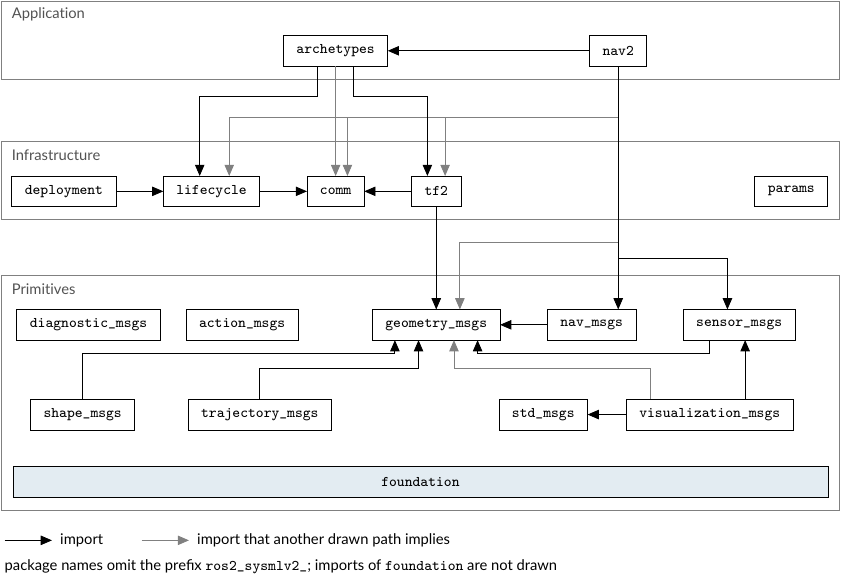}
\caption{The seventeen packages of \texttt{ros2-sysmlv2} in three layers, with the imports between them as declared in the package sources. An arrow points from the importing package to the imported one. A grey arrow is a declared import that a path of other drawn arrows already implies. Package names omit the prefix \texttt{ros2\_sysmlv2\_}, and imports of \texttt{foundation} are not drawn.}
\label{rsl:fig:layers}
\end{figure}

\texttt{ros2-sysmlv2} organizes its seventeen packages in three layers, shown with the imports between them in \cref{rsl:fig:layers}. The Primitives layer holds the foundation package and the message packages, the counterpart of the ROS2 \texttt{common\_\allowbreak interfaces} and \texttt{rcl\_\allowbreak interfaces} repositories. The Infrastructure layer models communication, QoS, lifecycle, deployment, parameters, and TF2, the concepts that the client libraries \texttt{rclpy} and \texttt{rclcpp}, the middleware, and the \texttt{tf2} library implement. The Application layer contains general node archetypes and showcases Nav2, the ROS2 navigation package, as the first specialized ROS2 package. No other library package imports it. Imports run within a layer or toward a lower one and form no cycle, so a downstream model that needs only message definitions imports nothing from the Nav2 package.

\subsection{Messages and Communication Endpoints}
\label{rsl:sec:messages-endpoints}

The Primitives layer represents ROS2 message definitions as SysML~v2 items. It carries message field names, nested composition, cardinalities, and a documented naming convention from ROS2 identifiers to SysML~v2 identifiers. Floating-point widths map to \texttt{Real}, integral widths map to \texttt{Integer}, strings map to \texttt{String}, and Boolean values map to \texttt{Boolean}. The level constants of \texttt{DiagnosticStatus} become an enumeration, and a field is renamed when its name is a SysML~v2 keyword, such as \texttt{message}, or the name of a feature that every item definition inherits, such as \texttt{start}. Fixed arrays retain their declared cardinality, and variable arrays use an unbounded cardinality. Array features are ordered, and arrays of primitive values are also nonunique, because a message may repeat a value. \Cref{rsl:lst:laserscan} shows the mapping for \texttt{sensor\_msgs/LaserScan}. The field names follow the SysML~v2 convention, \texttt{float32} becomes \texttt{Real}, the header becomes a composite item, and the two variable arrays become ordered, nonunique features. The mapping does not keep numeric width or signedness, and the ROS2 identity of a message stays recoverable from its package and name.

\begin{figure}[!t]
\centering
\begin{minipage}[t]{0.30\textwidth}
\begin{lstlisting}
std_msgs/Header header
float32 angle_min
float32 angle_max
float32 angle_increment
float32 time_increment
float32 scan_time
float32 range_min
float32 range_max
float32[] ranges
float32[] intensities
\end{lstlisting}
\end{minipage}\hfill
\begin{minipage}[t]{0.66\textwidth}
\begin{lstlisting}[language=SysMLv2,morekeywords={ordered,nonunique}]
item def LaserScan {
    item header : Header;
    attribute angleMin : Real;
    attribute angleMax : Real;
    attribute angleIncrement : Real;
    attribute timeIncrement : Real;
    attribute scanTime : Real;
    attribute rangeMin : Real;
    attribute rangeMax : Real;
    attribute ranges : Real[0..*] ordered nonunique;
    attribute intensities : Real[0..*] ordered nonunique;
}
\end{lstlisting}
\end{minipage}
\captionof{lstlisting}{The ROS2 message \texttt{sensor\_msgs/LaserScan} and its library definition. Left, the field lines of \texttt{LaserScan.msg} with their comments removed. Right, the \texttt{item def} in the library package for \texttt{sensor\_msgs}, without its doc comment.}
\label{rsl:lst:laserscan}
\end{figure}

The Infrastructure layer defines publisher and subscriber ports with directed item features, drawn as declared, with the topic connection and the QoS profile, in Section~S1 of the supplementary material. A publisher uses an \texttt{out item}; a subscriber uses an \texttt{in item}. Service and action server and client definitions provide explicit request-response and goal-feedback-result roles. Each endpoint kind has a port definition of its own, so a tool selects publishers, subscribers, servers, and clients by specialization alone, and a conformance check confirms that the two definitions of each pair differ only in direction. The subscriber is not written as the conjugate of the publisher, because a conjugated port definition specializes its original in KerML \citep{omg2025kerml}, and selection by specialization would then read a subscriber as a publisher.

Topic, service, and action bindings are interface definitions. A topic connection joins one publisher port to one subscriber port, so a topic with several subscribers is modeled as one connection for each publisher and subscriber pair. Each port usage binds its own topic name, message type, and QoS profile, as the node does in its code. The validator checks the roles of the two ends. The extractor of \cref{rsl:sec:generated-package} stops when the two ends of a binding carry different message, service, or action types, and it reports topic, service, or action names that differ after namespace resolution. The generator warns when the QoS profiles of the two ends break the rule that the library's \texttt{QoSCompatible} constraint states, and the generated monitor of \cref{rsl:sec:conformance} compares both ends of each connection on the running graph.

QoS profiles are modeled as attribute definitions with the fields and policy enumerations used by the ROS2 client libraries. The library includes the standard preset profiles and represents duration-valued fields through typed referential items. The preset \texttt{sensorDataQoS}, for example, binds keep-last history with depth 5, best-effort reliability, and volatile durability, the values of \texttt{rmw\_qos\_profile\_sensor\_data}. This design keeps a policy associated with its endpoint declaration, so a specialized node can make its reliability, durability, history, and related choices visible in the architecture model.

\subsection{Lifecycle, Parameters, and Frames}
\label{rsl:sec:lifecycle-parameters-frames}

\begin{figure}[!t]
\centering
\includegraphics{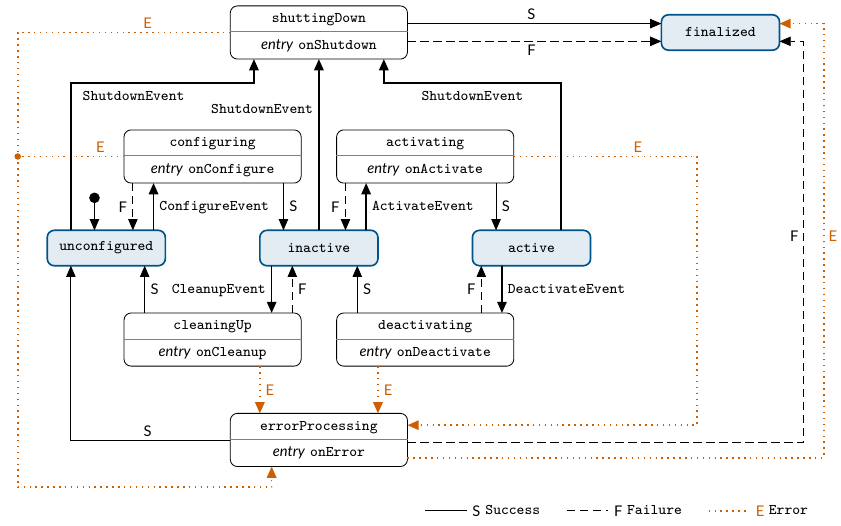}
\caption{The managed-node lifecycle as modeled in \texttt{LifecycleStates}, following the default state machine of \texttt{rcl\_lifecycle}. Shaded boxes are the four primary states. White boxes are the six transition states, each running its callback as an entry action. Thick arrows labeled with a request event are the seven requested transitions. Every transition state has three outcome transitions, guarded on its callback's result: Success (solid, S), Failure (dashed, F), and Error (dotted, E).}
\label{rsl:fig:lifecycle}
\end{figure}

Managed ROS2 nodes follow the default lifecycle state machine of \texttt{rcl}. The library models it as the state definition \texttt{LifecycleStates} in \cref{rsl:fig:lifecycle}, with the four primary states, the six transition states, and the 25 transitions that \texttt{rcl\_lifecycle} registers. A requested transition accepts a request event, such as \texttt{ConfigureEvent}, and leads from a primary state into a transition state, which runs its callback as its entry action and leaves by one of three outcome transitions, guarded on whether the callback returned \texttt{Success}, \texttt{Failure}, or \texttt{Error}. Section~S1 of the supplementary material lists both kinds of transition for the configuring state as declared. A \texttt{LifecycleNode} specializes the base \texttt{Node} definition and exhibits the lifecycle state machine. This separates node identity from managed-node behavior, so an unmanaged node can specialize the base definition without taking on lifecycle state.

Parameters are represented through parameter-type enumerations, integer and floating-point ranges, and descriptors that state a parameter's type, description, and admissible range, so a concrete architecture locates configuration choices with the node that owns them.

The TF2 vocabulary models a coordinate frame as a part with a frame identifier and models static and dynamic transforms as typed connections. Standard frames such as \texttt{map}, \texttt{odom}, and \texttt{base\_link} specialize the frame definition. A robot instance can compose these with sensor-mount transforms to make a frame tree explicit alongside its node and communication topology.

\subsection{Archetypes and the Nav2 Package}
\label{rsl:sec:archetypes-nav2}

The Application layer captures node roles that recur across ROS2 robot software. It defines eight abstract archetypes: \texttt{SensorDriver}, \texttt{Controller}, \texttt{Planner}, \texttt{Estimator}, \texttt{BehaviorCoordinator}, \texttt{MapProvider}, \texttt{PerceptionPipeline}, and \texttt{VelocityFilter}. Each archetype specializes \texttt{LifecycleNode} and declares a communication pattern. Concrete nodes then bind that pattern to message types, channel names, QoS presets, and parameter values through redefinition.

Nav2 is the first ROS2 package modeled in the library, and eight of its fourteen node definitions specialize an archetype. The \texttt{Controller} archetype and its specialization \texttt{ControllerServer}, drawn as declared in Section~S1 of the supplementary material, show the pattern. The specialization fixes the control frequency, gives the inherited action server, command publisher, and state subscriber their default names, and narrows their message items to \texttt{Twist} and \texttt{Odometry}. The same definition is in use in a hybrid ground-robot model that places the library's Nav2 stack beside its own sensor and localization subsystems and its mechanical, power, and compute parts. Topic connections whose ends are feature paths through the part hierarchy wire the nodes.

\subsection{Packaging and Downstream Use}
\label{rsl:sec:packaging}

The library is distributed as a KerML project archive, the interchange format that the language specification defines for reusable models \citep{omg2025kerml}, and published on the Sysand index \citep{sensmetry2026sysand} from a version tag of the public repository. The archive holds the seventeen source files, the changelog, the license, and two manifests that name the library, its version and its dependency on the KerML library and map each package name to its file. Versions follow semantic versioning, as the specification recommends, and the changelog marks the changes that require an existing model to be updated.

A robot model uses the library through package import and specialization. One command of the package manager records the library as a dependency in the model's own manifest and installs the sources, and the model then imports the packages it needs, so the robot's architecture stays its own model and the library enters it by reference. A second library builds on the first by the same route. We created a downstream package that declares the published release as its dependency and defines a monitoring node as a specialization of \texttt{LifecycleNode}; the validator resolves its imports from the installed copy and reports them as unresolved when the copy is absent. We expect domain extensions to take this route, for example a model of another ROS2 package or the shared vocabulary of a robot family. Their definitions specialize the library's definitions, so the projection of \cref{rsl:sec:selection} selects their usages as it selects usages of the library's own.

\subsection{Parity with the ROS2 Sources}
\label{rsl:sec:source-grounding}

The library is checked for parity with the ROS2~Jazzy sources. Parity means agreement under the mapping of \cref{rsl:sec:messages-endpoints}: a field name is compared after the naming convention is applied and a field type after it is mapped, so \texttt{float32} and \texttt{float64} both agree with \texttt{Real}. The message checker covers eight ROS2 message packages, in which the library models 88 of the 120 message types, every type of \texttt{geometry\_msgs}, \texttt{sensor\_msgs}, \texttt{nav\_msgs}, \texttt{diagnostic\_msgs}, \texttt{shape\_msgs}, and \texttt{trajectory\_msgs} among them. It compares the name, mapped type, cardinality, and array flags of every field, 311 fields in all. Further checkers compare the QoS enumerations, profile fields, and presets with \texttt{rclpy} and \texttt{rmw}, the lifecycle states, transitions, and callback results with \texttt{lifecycle\_msgs} and \texttt{rcl\_lifecycle}, the parameter messages with \texttt{rcl\_interfaces}, and the standard frame names with REP~105. For Nav2 they compare the 49 fields of the fourteen modeled action and message definitions with the \texttt{nav2\_msgs} sources, and they check that the server nodes are lifecycle nodes and that their core endpoints carry the modeled names. Run on the library as it stood before this comparison was added, the Nav2 field check reported four definitions that differed from the sources, and these were corrected. The sources are the Jazzy releases of \texttt{common\_interfaces} (5.3.7), \texttt{rcl\_interfaces} (2.0.3), \texttt{rcl} (9.2.11), \texttt{rclpy} (7.1.11), and \texttt{rmw} (7.3.3), and the Jazzy branch of \texttt{navigation2}. The library is in parity with the sources on every declaration checked.

\section{The Software Projection and the Generated Package}
\label{rsl:sec:projection}

\subsection{Selection by Type}
\label{rsl:sec:selection}

The type hierarchy lets a tool find the ROS2 elements of any model built on the library. Let $U$ be the set of usages in a model after its redefinitions and value bindings are resolved, each identified by its feature path from the system root. A usage $x$ can have several types, written $T(x)$, which KerML derives from its own typings and from the features it subsets or redefines \citep{omg2025kerml}. Let $R$ be the fourteen library definitions that head the structural kinds a ROS2 consumer reads: \texttt{Node}, \texttt{CoordinateFrame}, \texttt{DeclaredParameter}, the six port definitions, the three interface definitions, and the two transform connection definitions. The projection is the set of usages with a type that specializes one of them,
\begin{equation}
\pi_{\mathrm{ROS2}}(U) = \{\, x \in U : \exists\, t \in T(x),\ \exists\, r \in R,\ t \preceq r \,\},
\label{rsl:eq:projection}
\end{equation}
where $\preceq$ is specialization, reflexive and at any depth. A user's own subtype of a library port, interface, or connection definition is therefore selected, and a definition with the same name in another package is not. Message types, QoS profiles, and parameter values are not selected on their own. A tool reads them from the selected usages, a message type as the type of the item feature that a port carries and a QoS profile or a parameter value as an attribute value.

Selection keeps the hierarchy that connection ends rely on. A selected usage keeps its feature path as its identity, and a selected port belongs to the selected node that owns it. A usage that is not selected, such as the sensor suite that owns a sensor node in the hybrid model of \cref{rsl:sec:archetypes-nav2}, contributes a path segment and nothing else, so a connection end written as a path through such usages still resolves to a selected port.

The projection is well formed when every selected port or parameter is owned by a selected node and every end of a selected connection is a selected port, or a selected frame for a transform. The second condition follows from typing. Each library interface or connection definition types its ends by port or frame definitions in $R$, and the validator reports an end whose connected feature does not conform to that type. The first is a modeling rule that SysML~v2 does not impose, and a model that keeps it is the extractor's intended input. The bridge extractor of \cref{rsl:sec:generated-package} implements the selection for such models and writes only well-formed projections: it reads ports from selected nodes only, and it stops with an error when an end of a selected connection does not resolve to a selected port or frame. Whatever else the model holds is not selected and stays in place, as in \cref{rsl:fig:overview}.

The selection has four properties that the generated package and the monitor rely on, and we state them together.

\begin{proposition}[the projection]
\label{rsl:prop:projection}
Let $U$ be the usages of a model that passes the validator with warnings treated as errors, each with its feature path and its types, and let every selected port or parameter be owned by a selected node. Then $\pi_{\mathrm{ROS2}}(U)$ is the largest subset of $U$ whose usages all have a type that specializes a root in $R$. Every port and parameter in it is owned by a node in it, and every connection in it ends at a port or a frame in it, so it is a usage graph in its own right. It is determined by the model alone and not by the order in which a tool reads the model. If every usage has one feature path and every connection end resolves to one usage, a two-pass reference extractor computes $\pi_{\mathrm{ROS2}}(U)$ from the model's containment tree and stops on a model that breaks the ownership condition.
\end{proposition}

Section~S2 of the supplementary material proves the proposition, states the laws of the projection, among them that it is idempotent and distributes over union and intersection, and shows that the selection commutes with adding a second library or a downstream library's definitions to $R$. Section~S3 specifies the reference extractor and states what the bridge extractor of \cref{rsl:sec:generated-package} is checked against.

\subsection{The Generated Package}
\label{rsl:sec:generated-package}

\begin{figure}[!t]
\centering
\includegraphics{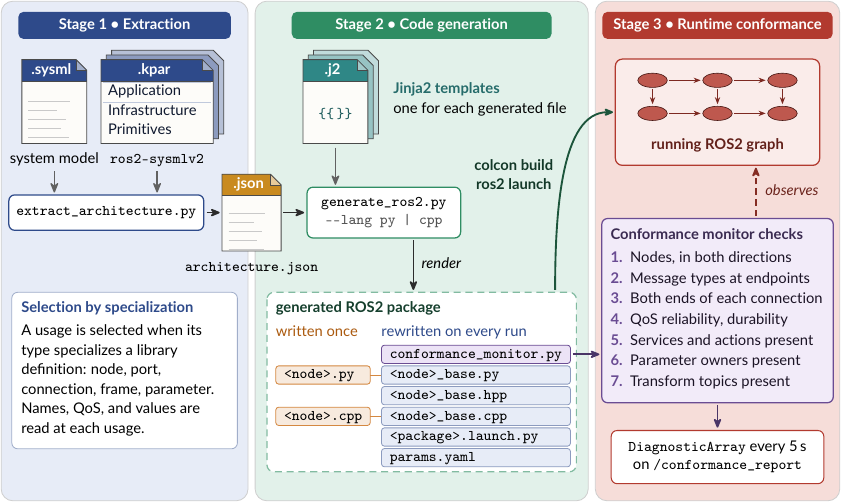}
\caption{The bridge pipeline in three stages. The extractor writes the records selected from the model and the library to \texttt{architecture.json}, and the generator renders a Python or C++ package from them. Files on a blue ground are rewritten on every run, and the node files on an orange ground are written once and then belong to the developer. The generated monitor (purple) is launched with the package, compares the running ROS2 graph with the extracted records, and publishes its report as ROS2 diagnostics.}
\label{rsl:fig:pipeline}
\end{figure}

The bridge pipeline of \cref{rsl:fig:pipeline} reads a model through the projection of \cref{rsl:eq:projection}. Its extractor reads node names, topic names, QoS, and parameter values in the scope of each usage, so a usage that redefines a default value of its definition contributes its own value, and it writes the selected records to an intermediate file. The generator renders a ROS2 package from these records in Python or C++. It checks all records before it writes the first file, and it stops when a port carries no message type, when a parameter has neither a known type nor a default value, or when a name is not a valid ROS2 name. One name resolver, following the name rules of \texttt{rcl}, gives the extractor, the launch file, the parameter file, and the monitor the same graph names. An endpoint's QoS becomes the client library's profile for the named preset, or a profile built from the model's values when the model writes out a profile of its own. \Cref{rsl:tab:stages} lists, for each ROS2 concept of the library, what the extractor reads, what the generator emits, and what the monitor of \cref{rsl:sec:conformance} checks on the running graph.

\begin{table}[!t]
\centering
\footnotesize
\setlength{\tabcolsep}{4pt}
\caption{What each stage of the bridge pipeline does with the ROS2 concepts of the library. A dash marks a stage that does not act on the concept in the version described here.}
\label{rsl:tab:stages}
\begin{tabular}{@{}>{\raggedright\arraybackslash}p{2.0cm}>{\raggedright\arraybackslash}p{3.0cm}>{\raggedright\arraybackslash}p{2.5cm}>{\raggedright\arraybackslash}p{3.1cm}>{\raggedright\arraybackslash}p{2.5cm}@{}}
\toprule
\textbf{ROS2 concept} & \textbf{In the library} & \textbf{Extracted} & \textbf{Generated} & \textbf{Checked on the running graph} \\
\midrule
Node & \texttt{Node}, \texttt{LifecycleNode}, eight archetypes & Name, namespace, archetype & Lifecycle base class, developer-owned file, launch entry & Presence, in both directions \\
Topic endpoint & Publisher and subscriber ports, message items & Topic name, message type & Publisher or subscriber, created in the configure callback & Message type at the owner's endpoint, connected or not \\
Topic connection & \texttt{TopicConnection} & Both ends as feature paths, message types and topic names compared & Warning on incompatible QoS profiles & Both ends present on the topic, QoS of the two ends compatible \\
QoS & Profile, policy enumerations, presets & Preset or profile with its policy values & Client-library profile of the preset or of the model's own profile & Reliability and durability at each publisher and connection end \\
Parameter & \texttt{DeclaredParameter}, descriptors, ranges & Name, type, default value & Declaration in the base class, parameter file keyed by full node name & Owner node present \\
Service and action & Server and client ports, bindings & Name and type, both compared across a binding & -- & Presence \\
Frame and transform & Frames, static and dynamic transforms & Frame names, parent and child & -- & Transform topics present \\
Lifecycle & State machine of \texttt{rcl} & Whether a node is managed & Base class that releases its endpoints on cleanup, shutdown, and error; lifecycle manager; activation script & -- \\
Deployment & Executors, containers, callback groups & -- & -- & -- \\
\bottomrule
\end{tabular}
\end{table}

For each node the generator emits a lifecycle base class. The base declares the modeled parameters when the node is created, creates the modeled publishers and subscribers in its configure callback, and destroys them on cleanup, on shutdown, and on an error, after which the node is unconfigured again. Every generated node is lifecycle-managed, as all eight archetypes are, and a generated script brings each node to the active state from the state it is in.

\begin{figure}[!t]
\centering
\begin{minipage}[t]{0.395\textwidth}
\begin{lstlisting}[language=SysMLv2]
part def LidarDriver :> SensorDriver {
    :>> nodeName = "lidar";
    :>> namespace = "/sensors";
    :>> updateRateHz = 10.0;
    :>> frameId = "lidar_link";
    port :>> sensorPub : TopicPublisher {
        :>> topicName = "/scan";
        :>> qos = sensorDataQoS;
        out item :>> msg : LaserScan;
    }
}
\end{lstlisting}
\end{minipage}\hfill
\begin{minipage}[t]{0.585\textwidth}
\begin{lstlisting}[language=Python]
from rclpy.qos import qos_profile_sensor_data
from sensor_msgs.msg import LaserScan

class LidarDriverBase(LifecycleNode):

    def __init__(self, node_name: str = 'lidar'):
        super().__init__(node_name=node_name)

    def on_configure(self, state: LifecycleState) -> TransitionCallbackReturn:
        self.sensorPub = self.create_lifecycle_publisher(
            LaserScan,
            '/scan',
            qos_profile_sensor_data)

        self.on_configure_hook()
        result = super().on_configure(state)
        return result
\end{lstlisting}
\end{minipage}
\captionof{lstlisting}{A node definition and the code generated from it. Left, the lidar driver of the ground-robot model. Right, an excerpt of its generated Python base class, without docstrings, logging calls, attribute initializations, and the other lifecycle callbacks.}
\label{rsl:lst:lidar}
\end{figure}

\Cref{rsl:lst:lidar} follows one node from the model to the code. The lidar driver of the ground-robot model specializes the \texttt{SensorDriver} archetype and binds its node name, namespace, topic, QoS preset, and message type. The generated base class creates that publisher in its configure callback, with the message class, the topic name, and the \texttt{rclpy} profile that corresponds to the preset. Because base files are rewritten on every run and derived files are not, a change to the model's interfaces reaches the code at the next run while the developer's algorithms stay in place. Section~S4 of the supplementary material describes the pipeline's stages, its wired mode for exercising a topology before the algorithms exist, and the release and dependency mechanics of the library in detail.

\section{Conformance of the Running Graph}
\label{rsl:sec:conformance}

Every five seconds the generated conformance monitor compares the running ROS2 graph with the extracted records and publishes its findings as ROS2 diagnostics. The last column of \cref{rsl:tab:stages} lists what it checks for each concept. It checks that every modeled node is present and no unmodeled one is, the message type at each modeled endpoint, and both ends of every modeled connection on their topic. It checks reliability and durability at each publisher and connection end against the declared preset, and the presence of the modeled services, actions, parameter owners, and transform topics. For each check it reports how many expectations agree among those it compared. An expectation without a concrete value, such as a QoS preset that leaves its policies to the middleware, is counted as not compared and never as agreeing. A missing node, a broken connection, or a wrong message type is reported at error level and a policy mismatch at warning level, each with the node and the topic named.

The comparison is defined on keyed records. The extractor normalizes each selected usage to a record with a key and coordinates. A node is keyed by its fully qualified name, a frame by its identifier, and a parameter by its owner and its name. An endpoint is keyed by its owner, its kind, and its resolved topic, service, or action name, with the message type as a coordinate, and a connection by the keys of its two ends. Every value that the model pins, such as the reliability of a named preset or a parameter default, is a coordinate at its key, and a value the model leaves open is not. The running graph is observed as records under the same keys, and the two record sets are compared kind by kind.

\begin{proposition}[structural conformance]
\label{rsl:prop:conformance}
Let the model's records and the observed records be keyed as above, with no two records of one kind under one key, and let the observation be complete for the kinds compared. The running graph conforms to the model exactly when, for each kind, the set of observed keys equals the set of modeled keys and every pinned coordinate agrees at its key. A departure is then one of three things, a modeled key with no observed record, an observed key with no modeled record, or a pinned coordinate whose two values differ, and it names its key and, for a value, its coordinate. Suppose the monitor samples the graph with gaps of at most $h$, obtains exact records once the graph has been constant for $\delta$, and publishes within $\eta$ of a sample. Then a departure that persists from its start through the first sample taken $\delta$ or more after it is reported within $\delta + h + \eta$ of its start.
\end{proposition}

The generated monitor is an instance of this comparison on the keys and coordinates that \cref{rsl:tab:stages} lists: nodes in both directions, the message type at each modeled endpoint, both ends of each modeled connection, and reliability and durability where the preset pins them. For services, actions, parameters, and transforms it compares presence. Section~S5 of the supplementary material gives the monitor's keys and counting rules. Section~S6 proves the proposition and develops the verdicts of a monitor that sees the graph through samples with bounds rather than exact records. It also states when a generated package carries the model's keys and pinned values to the running graph, so that a graph realized from the package conforms to the model.

\section{Case Study: Conformance Monitoring of a Generated Robot}
\label{rsl:sec:case-study}

\begin{figure}[!t]
\centering
\includegraphics{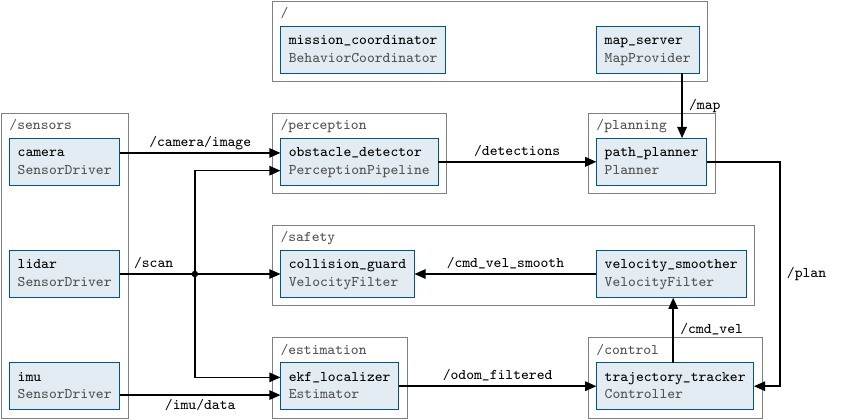}
\caption{Topic graph of the package generated from the ab initio ground-robot model, drawn from the package's \texttt{architecture.json}. Each box is a node, labeled with its node name and the archetype it specializes, and the grey frames group the nodes by namespace. Each arrow is a modeled topic connection, labeled with its topic.}
\label{rsl:fig:generated-graph}
\end{figure}

We apply the library to an ab initio autonomous-ground-robot (AGR) architecture. The model combines sensor, perception, estimation, planning, control, and safety nodes with the ROS2 definitions they specialize, and it does not use the Nav2 package. The same model holds the robot's structure and power system, its behavior, ten requirement definitions, and the allocation of all eleven nodes to the onboard computer, which the projection leaves in place. \Cref{rsl:fig:generated-graph} shows the topic graph of the package generated from it, with eleven nodes joined by eleven topic connections. The hybrid model of \cref{rsl:sec:archetypes-nav2} composes custom sensor, perception, and estimation nodes with the library's Nav2 stack. The same vocabulary thus serves a robot built entirely from its own nodes and one built around Nav2.

We built the generated Python package of the AGR model on ROS2~Jazzy and ran it with no simulator. Every publisher sent default-constructed messages on a timer, and the lifecycle manager of the generated launch file brought the nodes to the active state. In each of the 12 monitor cycles of the run, all 11 modeled nodes and all 11 modeled connections were present, and the message types at the 22 ends of these connections agreed with the model. Ten of the model's eleven publishers declare a QoS preset. Reliability and durability agreed on the 8 topics whose preset fixes a concrete value, and the 2 whose preset defers to the middleware were counted as not compared. The monitor reported the modeled map service, the three modeled actions, and the transform topics as absent, at warning level. Generating these interfaces from the extracted records is the next template step.

\begin{table}[!t]
\centering
\footnotesize
\caption{Changes to the lidar driver's developer-owned file and the monitor's report. \Cref{rsl:lst:lidar} shows what the model declares.}
\label{rsl:tab:deviations}
\begin{tabular}{@{}p{4.6cm}p{7.6cm}p{1.2cm}@{}}
\toprule
\textbf{Change in the lidar driver} & \textbf{Monitor report} & \textbf{Level} \\
\midrule
None (control) & Same findings as the unchanged package & \\
Scan published on \texttt{/scan\_raw} & Three scan connections broken, lidar publisher absent from \texttt{/scan} & Error \\
\texttt{String} messages on \texttt{/scan} & Type mismatch at the lidar publisher, \texttt{LaserScan} expected & Error \\
Reliable, transient-local delivery on \texttt{/scan} & Policy mismatch on \texttt{/scan} and at the publisher end of its three connections & Warning \\
\bottomrule
\end{tabular}
\end{table}

We then let the implementation depart from the model. We changed only the developer-owned file of the lidar driver of \cref{rsl:lst:lidar}, with the model and the monitor unchanged, and ran the package once per change. \Cref{rsl:tab:deviations} lists the changes and what the monitor reported. It reported each change from its first cycle and named the node and the topic, as \cref{rsl:prop:conformance} says a departure is named, by the publisher's key and, for the changed policy, by the reliability and durability coordinates at that key. An engineer thus learns of a departure from the architecture while the system runs. Section~S7 of the supplementary material holds the record of these runs, with the monitor's lines for the baseline and for each change in both languages.

\section{Conclusion}
\label{rsl:sec:conclusion}

We presented \texttt{ros2-sysmlv2}, a native SysML~v2 domain library for modeling ROS2 robotic systems, and the round trip that a model built on it supports. A robot model specializes the library's definitions to declare nodes, ports, connections, frames, and parameters, and a tool selects these elements by their library types. That selection is the largest set of usages with a library type, it is determined by the model alone, and a ROS2 package and a conformance monitor are generated from it. On every declaration checked, the library agrees with the ROS2~Jazzy sources under its naming and type mapping. Two ground-robot models were built on it, and the monitor generated from one of them reported each departure we introduced in the implementation, named by the key and the value at which the running graph differed from the model. Each model built on the library declares its software interfaces against the same definitions, so a publisher, a binding, or a frame carries the same meaning in every robot model and is read by the same tools. The library thus gives systems engineers a reusable foundation for bringing ROS2 architecture into a native SysML~v2 model, and the generated monitor reports where the running system departs from that model.

\backmatter

\section*{Significance and Practitioner Points}
For practitioners, a systems engineer who models a ROS2 robot in SysML~v2 can start at once. One package-manager command installs the published library, and the robot model then declares its nodes, ports, connections, frames, and parameters by specializing library definitions, with no need to model ROS2 first. The library covers core ROS2 and Nav2, the ROS2 navigation package, and further ROS2 packages are added as libraries that build on it by the same route. For researchers, this article shows that the architectural concepts of ROS2 can be stated with the definition and usage vocabulary of SysML~v2 itself, without a profile or a separate metamodel, and that such a library can be held to the software it models. Its declarations are compared with the ROS2~Jazzy sources, and the comparison found four definitions that differed and were then corrected. Because every ROS2 element of a model specializes a library definition, a tool finds those elements by type, and we use this selection to generate ROS2 package skeletons and a monitor that reports where the running system departs from the architecture.

\section*{Acknowledgements}
The authors thank Sensmetry for providing academic licenses for the Syside Automator and Modeler tools.

\section*{Conflict of Interest}
The authors declare no conflict of interest.

\section*{Data Availability Statement}
The \texttt{ros2-sysmlv2} library that this article describes is openly available under the Apache~2.0 license. Its sources are in the public repository at \url{https://github.com/SEIL-ISR/ros2-sysmlv2}, and its releases are published as packages on the Sysand index at \url{https://sysand.com/projects/seil-isr/ros2-sysmlv2/}. The ROS2~Jazzy sources with which the library is compared are public.

\section*{Supporting Information}
The supplementary material has seven sections. Section S1 gives the library in more detail: the ROS2 concepts it models, its endpoint, QoS, and lifecycle declarations as written in the sources, the path from an archetype to a robot model, its archive and release mechanics, and the parity check in full. Sections S2 and S3 give the definitions and proofs behind the projection of \cref{rsl:sec:projection} and specify the reference extractor against which the bridge extractor is checked. Sections S4 and S5 describe the bridge pipeline and the generated monitor in detail. Section S6 gives the structural conformance results behind \cref{rsl:sec:conformance} with their proofs, the verdicts of a monitor that observes the graph through samples, and the conditions under which a generated package carries the model's keys to the running graph. Section S7 is the record of the case study of \cref{rsl:sec:case-study}.

\bibliography{bibliography/references}

\clearpage
\setcounter{section}{0}
\setcounter{equation}{0}
\setcounter{table}{0}
\setcounter{figure}{0}
\setcounter{lstlisting}{0}
\setcounter{theorem}{0}
\setcounter{footnote}{0}
\renewcommand{\thetheorem}{S\arabic{theorem}}
\renewcommand{\thesection}{S\arabic{section}}
\renewcommand{\theequation}{S\arabic{equation}}
\renewcommand{\thetable}{S\arabic{table}}
\renewcommand{\thefigure}{S\arabic{figure}}
\renewcommand{\thelstlisting}{S\arabic{lstlisting}}
\renewcommand{\theHsection}{S\arabic{section}}
\renewcommand{\theHequation}{S\arabic{equation}}
\makeatletter
\titleformat{\section}
  {\fontsize{10pt}{13\p@}\bfseries}
  {\thesection}{1em}{\textmd{|}\quad\allcaps{#1}}
\makeatother
\titlespacing*{\section}{0pt}{1.5\baselineskip}{\baselineskip}

\section*{Supplementary Material for\\ A SysML~v2 Domain Library for ROS~2 Robot Architectures}

This material accompanies the article and has seven sections, in the order of the article. \Cref{rsls:sec:library} gives the library in more detail than \cref{rsl:sec:library} of the article. It holds the ROS2 concepts the library models, its endpoint, QoS and lifecycle declarations as written in the sources, the path from an archetype to a robot model, its archive and release mechanics, and the parity check against the ROS2~Jazzy sources in full. \Cref{rsls:sec:projection} defines the projection of \cref{rsl:eq:projection} of the article on the usages of a model, proves \cref{rsl:prop:projection} of the article, and states how the selection behaves under a second library and a downstream library. \Cref{rsls:sec:extractor} specifies the reference extractor that computes the projection from the containment tree and states what the bridge extractor is checked against. \Cref{rsls:sec:pipeline} follows the bridge pipeline from the model file to the launched package, and \cref{rsls:sec:monitor} gives the keys, diagnostic groups and counting rules of the generated monitor. \Cref{rsls:sec:conformance} defines keyed conformance of the running graph and proves \cref{rsl:prop:conformance} of the article. It gives the verdict of a monitor that observes the graph through samples, the detection bound, and the conditions under which a generated package carries the model's records to the running graph. \Cref{rsls:sec:case-record} is the record of the case study of \cref{rsl:sec:case-study} of the article. It holds the generated graph's records, the raw lines of the monitor for the baseline run and for each of the three departures, and the runs of the same package in Python and in C++. Sections, equations, propositions, figures, tables and listings of the article are named by their number there; those of this material carry the prefix S.

\section{The Library in Detail}
\label{rsls:sec:library}

This section carries the parts of the library's description that \cref{rsl:sec:library} of the article gives in condensed form. They are the ROS2 concepts the library models, and its endpoint and QoS declarations and its lifecycle state machine as the sources declare them. They are also the path from an archetype to a robot model, the archive and release mechanics of the published package, and the parity check against the ROS2~Jazzy sources in full.

\subsection{ROS2 Concepts}
\label{rsls:sec:ros2-concepts}

A ROS2 system is a set of \emph{nodes}. A node is a named unit of computation, usually one process or one component within a process, and it lives in a namespace that qualifies its name. Nodes discover one another at run time, and the nodes together with the communication among them form the ROS2 graph. The graph is the software architecture of the robot as it runs.

Nodes exchange data in three ways. A \emph{topic} is a named channel that carries messages of one type. Any number of nodes may publish on it and any number may subscribe, and a publisher and a subscriber are matched by the topic name and the message type. A \emph{service} is a request answered by a response, offered by one server node and called by client nodes. An \emph{action} is a goal that takes time to complete. A client sends the goal, and the server reports feedback while it works and a result at the end. Message, service, and action types are declared in interface files, and standard packages such as \texttt{sensor\_msgs} define the types that most robots share.

Every publisher and subscriber carries QoS policies. Reliability states whether a lost message is sent again (reliable) or dropped (best effort). Durability states whether a subscriber that joins late receives the messages already published (transient local) or only later ones (volatile). History and depth bound the message queue, and deadline, lifespan, and liveliness set timing expectations. A publisher and a subscriber are connected only when the publisher offers at least what the subscriber requests, so a reliable subscriber receives nothing from a best-effort publisher. ROS2 names common combinations as presets, such as the sensor-data preset for high-rate sensor streams.

A node is configured through parameters, which are named, typed values that the node declares and that a launch file or a parameter file sets at start-up. A managed node also follows a standard lifecycle. It starts unconfigured, becomes inactive once configured, and processes data only when active. An external manager requests each transition, and the node runs a callback whose result decides whether the transition succeeds. The lifecycle lets a system bring its nodes up in a known order and confirm that each is configured before data flows.

Coordinate frames are handled by the \texttt{tf2} library. Nodes broadcast the transform between a parent frame and a child frame, and the transforms form a tree that any node can query. ROS Enhancement Proposal (REP)~105 fixes the names of the standard frames of a mobile robot, among them \texttt{map}, \texttt{odom}, and \texttt{base\_link}. Application code reaches all of these concepts through a client library, \texttt{rclcpp} for C++ or \texttt{rclpy} for Python. Both are built on the common library \texttt{rcl}, which calls the communication middleware through the \texttt{rmw} interface. ROS2 is released in distributions, and the article uses Jazzy, the long-term-support distribution of 2024. Nav2 is the navigation package of ROS2 \citep{macenski2020marathon}. It consists of managed server nodes for planning, control, and recovery behaviors, which a navigator node coordinates through actions.

\subsection{Endpoints, Topic Connections, and QoS Profiles as Declared}
\label{rsls:sec:endpoints-declared}

\begin{figure}[!t]
\centering
\includegraphics{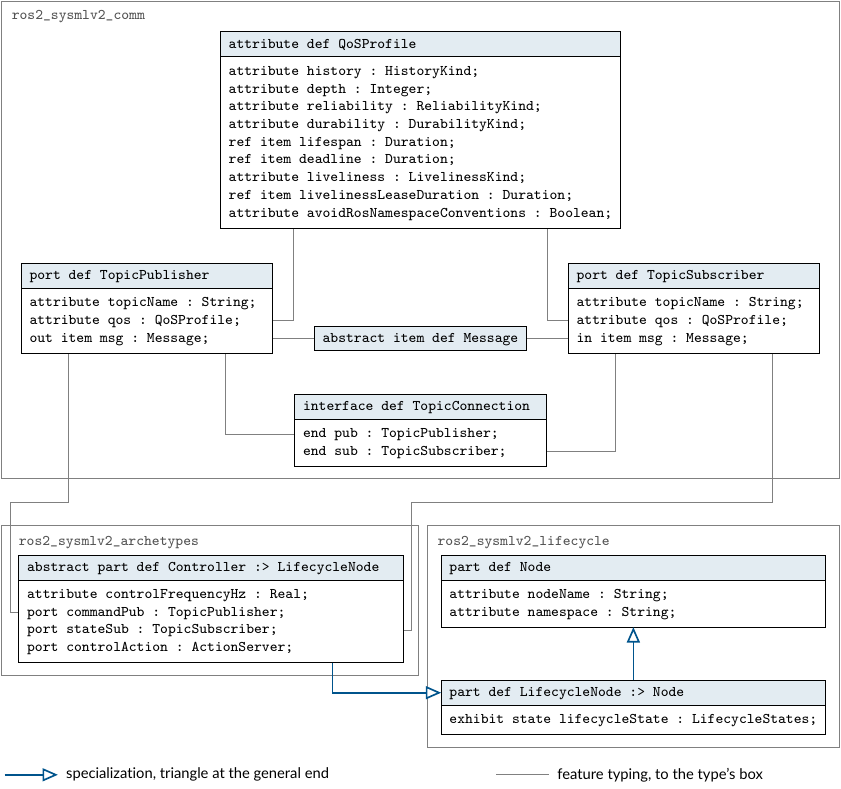}
\caption{A slice of the library vocabulary in three packages, as declared in the sources. Each box is a definition with its features. Arrows with a hollow head are specializations and point to the general definition. Thin lines join a feature to the box of its type. The publisher and subscriber ports each carry a topic name, a QoS profile, and a message item, directed out of the publisher and into the subscriber, and the topic connection relates the two through its ends. The QoS profile is an attribute definition, so its duration fields take the \texttt{ref item} form. The \texttt{Controller} archetype obtains the lifecycle state machine by specializing \texttt{LifecycleNode}.}
\label{rsls:fig:vocabulary}
\end{figure}

The Infrastructure layer defines publisher and subscriber ports with directed item features, shown with the topic connection and the QoS profile in \cref{rsls:fig:vocabulary} as the sources declare them. Each port carries a topic name, a QoS profile, and a message item, directed out of the publisher and into the subscriber, and the topic connection relates the two ports through its ends. The two port definitions differ only in that direction, which is what lets a tool select publishers and subscribers by specialization alone, as \cref{rsl:sec:messages-endpoints} of the article explains. The QoS profile is an attribute definition, so its duration fields take the \texttt{ref item} form of \cref{rsl:sec:semantic-discipline} of the article, and the \texttt{Controller} archetype obtains the lifecycle state machine by specializing \texttt{LifecycleNode}.

\subsection{The Lifecycle State Machine as Declared}
\label{rsls:sec:lifecycle-declared}

The state definition \texttt{LifecycleStates} of \cref{rsl:fig:lifecycle} of the article has the four primary states, the six transition states, and the 25 transitions that \texttt{rcl\_lifecycle} registers. Seven of the transitions are requested ones. Each accepts a request event, such as \texttt{ConfigureEvent}, and leads from a primary state into a transition state. The other eighteen are outcome transitions. A transition state runs its callback as its entry action and leaves by one of three transitions, guarded on whether the callback returned \texttt{Success}, \texttt{Failure}, or \texttt{Error}. \Cref{rsls:lst:lifecycle} shows both kinds of transition for the configuring state.

\begin{figure}[!t]
\begin{lstlisting}[language=SysMLv2,morekeywords={entry,if}]
action def OnConfigure {
    out attribute result : TransitionCallbackReturn;
}

state configuring {
    entry action onConfigure : OnConfigure;
}

transition configure
    first unconfigured
    accept ConfigureEvent
    then configuring;

transition onConfigureSuccess
    first configuring
    if configuring.onConfigure.result == TransitionCallbackReturn::Success
    then inactive;
\end{lstlisting}
\captionof{lstlisting}{Excerpt of \texttt{LifecycleStates} for the configuring state, without doc comments. From the top, the callback action and its result, the transition state that runs it on entry, the requested transition into that state, and its outcome transition on success.}
\label{rsls:lst:lifecycle}
\end{figure}

\subsection{From Archetype to Robot Model}
\label{rsls:sec:archetype-to-robot}

\begin{figure}[!t]
\centering
\includegraphics{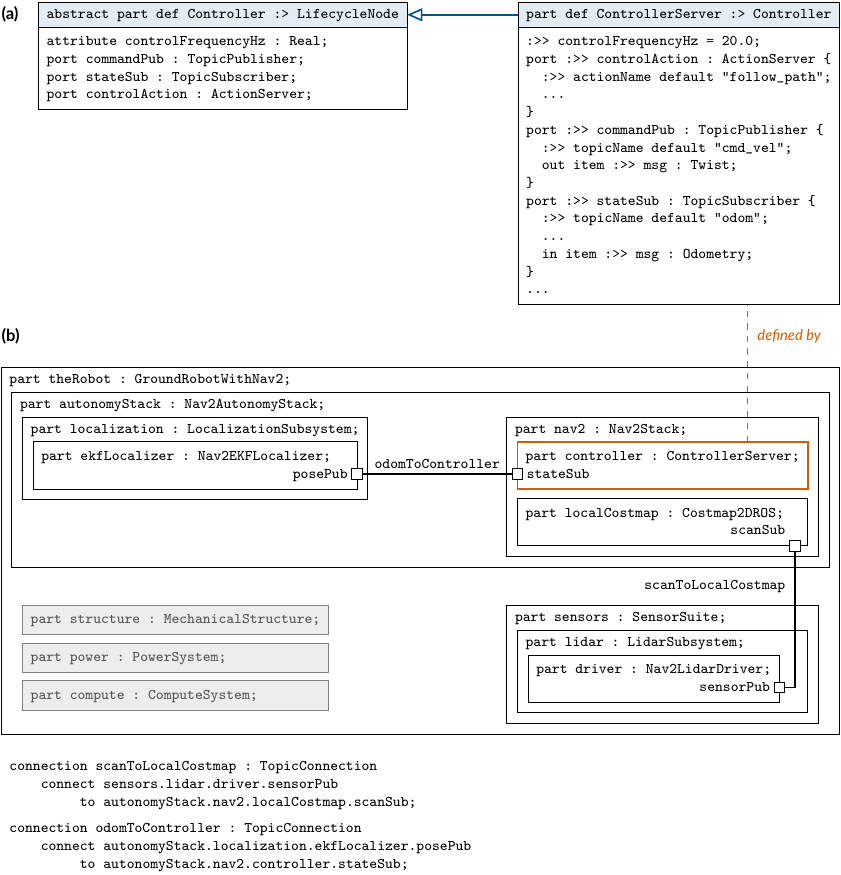}
\caption{From archetype to robot model. (a) The library archetype \texttt{Controller} and its Nav2 specialization \texttt{ControllerServer}, as declared. An ellipsis marks lines left out. (b) The usage of \texttt{ControllerServer} (orange box) inside the library's Nav2 stack, in an excerpt of a hybrid ground-robot model. Grey boxes are the robot's mechanical, power, and compute parts. The two topic connections, printed below the panel as declared, join ports through feature paths.}
\label{rsls:fig:archetype-to-robot}
\end{figure}

\Cref{rsls:fig:archetype-to-robot}(a) shows the \texttt{Controller} archetype and its specialization \texttt{ControllerServer}. The specialization fixes the control frequency, gives the inherited action server, command publisher, and state subscriber their default names, and narrows their message items to \texttt{Twist} and \texttt{Odometry}. Panel (b) shows the same definition in use. A hybrid ground-robot model places the library's Nav2 stack beside its own sensor and localization subsystems and its mechanical, power, and compute parts, and it wires the nodes with topic connections whose ends are feature paths through the part hierarchy. The reusable pattern and its binding in a concrete robot thus sit in one type hierarchy. The projection of \cref{rsls:sec:projection} selects the server node inside the Nav2 stack through the path that the stack and its subsystems contribute, and \cref{rsls:sec:extractor} states how an extractor reads such a path.

\subsection{Packaging, Release, and Downstream Use}
\label{rsls:sec:packaging-detail}

The library is distributed as a KerML project archive, the interchange format that the language specification defines for reusable models \citep{omg2025kerml}. The archive holds the seventeen source files, the changelog, the license, and two manifests. The first manifest names the library, its publisher, its version, and its license, and declares the KerML library on which the sources depend. The second maps each of the seventeen package names to its file, so that a tool resolves an import by package name. A release is built from a version tag of the public repository by an automated workflow, which compares the tag with the declared version and the archive with its expected file list, and then publishes the archive on the Sysand index \citep{sensmetry2026sysand}. Versions follow semantic versioning, as the specification recommends, and the changelog marks the changes that require an existing model to be updated.

A robot model uses the library through package import and specialization. One command of the package manager records the library as a dependency in the model's own manifest and installs the sources in the model's project, and the model then imports the packages it needs. The robot's architecture therefore stays its own model, and the library enters it by reference. As a concrete case, we built a small robot model on the library. A lidar driver and a controller each specialize a library archetype and bind their own topic names, payload types, and QoS preset. The robot definition composes these two nodes with the library's Nav2 stack and standard frame tree, and wires them with a topic connection and a static transform. The model passes the validator together with the library sources.

A second library builds on the first by the same route. We created a downstream package that declares the published release as its dependency and defines a monitoring node as a specialization of \texttt{LifecycleNode}. The package manager installs the release and records its version in a lock file. The validator then resolves the imports of the downstream package from the installed copy, and it reports them as unresolved when the copy is absent. We expect domain extensions to take this route, for example a model of another ROS2 package or the shared vocabulary of a robot family. Their definitions specialize the library's definitions, so the projection of \cref{rsl:eq:projection} of the article selects their usages as it selects usages of the library's own, which is the second claim of \cref{rsls:prop:libraries}.

\subsection{The Parity Check}
\label{rsls:sec:parity-detail}

The library is checked for parity with the ROS2~Jazzy sources. Parity means agreement under the mapping of \cref{rsl:sec:messages-endpoints} of the article. A field name is compared after the naming convention is applied and a field type after it is mapped, so \texttt{float32} and \texttt{float64} both agree with \texttt{Real}. A checker reads a ROS2 source and the library declaration that models it, and the two must agree. The message checker covers eight ROS2 message packages, in which the library models 88 of the 120 message types. These are every type of \texttt{geometry\_msgs}, \texttt{sensor\_msgs}, \texttt{nav\_msgs}, \texttt{diagnostic\_msgs}, \texttt{shape\_msgs}, and \texttt{trajectory\_msgs}, six of the thirty in \texttt{std\_msgs}, and four of the twelve in \texttt{visualization\_msgs}. For each one the checker reads the \texttt{.msg} file and compares the name, mapped type, cardinality, and array flags of every field, 311 fields in all. Further checkers compare the QoS enumerations, profile fields, and presets with \texttt{rclpy} and \texttt{rmw}, and the lifecycle states, transitions, and callback results with \texttt{lifecycle\_msgs} and \texttt{rcl\_lifecycle}. They also compare the parameter messages with \texttt{rcl\_interfaces} and the standard frame names with REP~105. For Nav2 they compare the 49 fields of the fourteen modeled action and message definitions with the \texttt{nav2\_msgs} sources in the same way, and they check that the server nodes are lifecycle nodes and that their core endpoints carry the modeled names. Run on the library as it stood before this comparison was added, the Nav2 field check reported four definitions that differed from the sources, and these were corrected. The sources are the Jazzy releases of \texttt{common\_interfaces} (5.3.7), \texttt{rcl\_interfaces} (2.0.3), \texttt{rcl} (9.2.11), \texttt{rclpy} (7.1.11), and \texttt{rmw} (7.3.3), and the Jazzy branch of \texttt{navigation2}. The library is in parity with the sources on every declaration checked. A model built on the library therefore starts from declarations that were compared with the ROS2 sources.

\section{The Projection and Its Laws}
\label{rsls:sec:projection}

The article selects the ROS2 elements of a model by type, as the projection of its \cref{rsl:eq:projection}. This section makes the objects of that definition explicit, proves \cref{rsl:prop:projection} of the article, and states how the selection behaves when a model is built on two libraries or on a library that extends \texttt{ros2-sysmlv2}. The usage graph below is the evaluated structural view that a consumer reads. It is not a metamodel of SysML~v2.

\begin{definition}[usage graph]
\label{rsls:def:graph}
Let $U$ be the set of usages of a model after its redefinitions and value bindings are resolved, as in the article, together with the system root, which has the empty feature path and the system definition as its type. Each usage $x \in U$ has a feature path $q(x)$ from the system root, a finite nonempty set of types $T(x) \subseteq \mathcal{T}$, and the values bound in its scope. Here $(\mathcal{T}, \preceq)$ is the set of definitions of the model and its libraries, ordered by specialization, reflexive and at any depth. Three kinds of usage carry structure. Let $P \subseteq U$ be the port and parameter usages and $E \subseteq U$ the connection usages. The owner map $\mathrm{own} \colon P \to U$ sends a port or parameter to the usage whose feature path is its own path with the last segment removed. The end maps $\partial_0, \partial_1 \colon E \to U$ send a connection to the two usages that its ends resolve to through their feature paths. A subset $V \subseteq U$ is \emph{incidence closed} when $\mathrm{own}(p) \in V$ for every $p \in V \cap P$ and $\partial_0(e), \partial_1(e) \in V$ for every $e \in V \cap E$.
\end{definition}

The maps restricted to an incidence closed subset take their values in it, so such a subset with the restricted maps, types and values is a usage graph in its own right. The article's owner condition covers ports and parameters, and the bridge extractor reads both by the same rule, so $P$ holds both.

For a set $\Lambda \subseteq \mathcal{T}$ of root definitions, let
\begin{equation}
\mathord{\downarrow}\Lambda = \{\, t \in \mathcal{T} : \exists\, r \in \Lambda,\ t \preceq r \,\}, \qquad \sigma_\Lambda(x) = \mathbf{1}\{\, T(x) \cap \mathord{\downarrow}\Lambda \neq \varnothing \,\}, \qquad \pi_\Lambda(V) = \{\, x \in V : \sigma_\Lambda(x) = 1 \,\}
\label{rsls:eq:selection}
\end{equation}
for $V \subseteq U$. With $\Lambda = R$, the fourteen root definitions of the article, $\pi_R$ is the projection $\pi_{\mathrm{ROS2}}$ of its \cref{rsl:eq:projection}. The predicate tests specialization and not the spelling of a name, and it admits several types on one usage, which KerML derives from the usage's own typings and from the features it subsets or redefines \citep{omg2025kerml}.

A connection usage has an identity of its own, so two connections may have the same two ends. Selecting a connection tests the connection's own types, even when both of its ends are already selected.

\begin{definition}[admitted selection]
\label{rsls:def:admitted}
The selection $\sigma_\Lambda$ is incidence closed on $U$ when every selected port or parameter has a selected owner and every selected connection has two selected ends, that is, when
\begin{equation}
\sigma_\Lambda(p) = 1 \ \Longrightarrow\ \sigma_\Lambda(\mathrm{own}(p)) = 1 \quad (p \in P), \qquad \sigma_\Lambda(e) = 1 \ \Longrightarrow\ \sigma_\Lambda(\partial_0(e)) = \sigma_\Lambda(\partial_1(e)) = 1 \quad (e \in E).
\label{rsls:eq:closure}
\end{equation}
\end{definition}

These are the two well-formedness conditions of \cref{rsl:sec:selection} of the article, with the role of each selected usage left to the lemma below. Membership of a port's type in the library does not by itself make the port's owner a selected usage, so the first condition is a condition on the model. The second condition holds in every validated model built on the library.

The interpretation rejects a selection that breaks either condition. Rejecting a dangling selected usage differs from silently pruning it, because pruning changes the specified selection.

\begin{lemma}[end typing of the library's connections]
\label{rsls:lem:ends}
Let a model pass the validator, with warnings treated as errors, together with the library sources, and let $e \in E$ be a connection usage with $\sigma_R(e) = 1$. Then both ends of $e$ are selected. If a type of $e$ specializes the topic connection, the service binding or the action binding definition, the two ends are port usages whose types specialize the two port definitions that the ends of that definition name, in the roles those ends name. If a type of $e$ specializes the static or the dynamic transform definition, the two ends are part usages whose types specialize the coordinate frame definition.
\end{lemma}

\begin{proof}
Each of the three interface definitions of the library types its two ends by two of the six port definitions in $R$. The topic connection names the publisher and the subscriber port, and the service binding and the action binding each name their server and client ports. Each of the two transform connection definitions types its two ends by the coordinate frame definition in $R$. A usage typed by one of these definitions, or by a definition that specializes one, inherits the two ends, and an end that is redefined keeps the redefined end's type among its types, as $T(x)$ records. The connected feature of an end must conform to the end's type, and the validator reports an end whose connected feature does not. So each end of $e$ resolves to a usage with a type at or below a port definition or the frame definition in $R$, which $\sigma_R$ selects, and that usage carries the role of the end that reaches it.
\end{proof}

For a validated model the second condition of \cref{rsls:eq:closure} therefore holds with $\Lambda = R$, and incidence closure reduces to the owner condition, that every port or parameter with a library type sits on a usage with a library type, in practice a node. A model that keeps this rule is the extractor's intended input. How the extractor treats a model that breaks it is stated in \cref{rsls:sec:extractor}.

\begin{proposition}[projection laws and maximality]
\label{rsls:prop:laws}
Let $\sigma_\Lambda$ be incidence closed on $U$, and let $V, V_1, V_2 \subseteq U$ be incidence closed subsets. Then the following hold.
\begin{enumerate}
\item $\pi_\Lambda(V)$ is incidence closed, hence a usage graph with the restricted maps, types and values.
\item $\pi_\Lambda(\pi_\Lambda(V)) = \pi_\Lambda(V)$.
\item $V_1 \subseteq V_2$ implies $\pi_\Lambda(V_1) \subseteq \pi_\Lambda(V_2)$.
\item $\pi_\Lambda(V_1 \cup V_2) = \pi_\Lambda(V_1) \cup \pi_\Lambda(V_2)$ and $\pi_\Lambda(V_1 \cap V_2) = \pi_\Lambda(V_1) \cap \pi_\Lambda(V_2)$.
\item $\pi_\Lambda(V)$ is the largest subset of $V$ all of whose usages satisfy $\sigma_\Lambda$, and it is determined by $V$ alone.
\end{enumerate}
\end{proposition}

\begin{proof}
Let $S = \{\, x \in U : \sigma_\Lambda(x) = 1 \,\}$, a fixed subset of $U$. Then $\pi_\Lambda(V) = V \cap S$ for every $V \subseteq U$. For the first claim, a port or parameter $p \in V \cap S$ has $\mathrm{own}(p) \in V$ because $V$ is incidence closed, and $\mathrm{own}(p) \in S$ because $\sigma_\Lambda$ is incidence closed on $U$. The same argument places both ends of a connection of $V \cap S$ in $V \cap S$. The second, third and fourth claims are the laws of intersection with the fixed set $S$. Intersecting twice with $S$ is intersecting once, inclusion is preserved by intersection, and intersection distributes over union and over intersection. For the last claim, every subset of $V$ whose usages satisfy $\sigma_\Lambda$ is a subset of $S$ and hence of $V \cap S$, and $V \cap S$ is itself such a subset. The set is defined pointwise, so two readings of the same evaluated model select the same usages in whatever order they visit them.
\end{proof}

\Cref{rsl:prop:projection} of the article follows with $\Lambda = R$ and $V = U$. Its model passes the validator with warnings treated as errors, so by \cref{rsls:lem:ends} every selected connection has two selected ends. Its owner condition is the first condition of \cref{rsls:eq:closure}. Together they make $\sigma_R$ incidence closed on $U$, and $U$ itself is incidence closed. The fifth claim then gives that $\pi_R(U)$ is the largest subset of $U$ whose usages have a type under a root in $R$ and that it is determined by the model alone. The first claim gives that it is a usage graph whose ports and parameters are owned inside it and whose connections end inside it. The extractor clause is \cref{rsls:thm:extraction} of \cref{rsls:sec:extractor}.

\begin{remark}[what the laws do not say]
\label{rsls:rem:laws}
The laws hold for subsets of one evaluated model. They do not say that two fragments of a model evaluated separately keep the references that cross between them, and they do not say that two models placed side by side gain or lose connections between their ports. Those are questions about evaluation and about the library's connection definitions, not about the selection. Types and values are inherited from the one evaluated model whose subsets are compared, and re-evaluating a changed source is a separate operation outside these laws.
\end{remark}

\begin{proposition}[a second library and a downstream library]
\label{rsls:prop:libraries}
Let $\Lambda_1, \Lambda_2 \subseteq \mathcal{T}$ be root sets whose selections are incidence closed on $U$, and write $D_i = \mathord{\downarrow}\Lambda_i$. Then the following hold for every $V \subseteq U$.
\begin{enumerate}
\item $\pi_{\Lambda_1 \cup \Lambda_2}(V) = \pi_{\Lambda_1}(V) \cup \pi_{\Lambda_2}(V)$, and the selection of $\Lambda_1 \cup \Lambda_2$ is incidence closed on $U$.
\item $\pi_{D_1}(V) = \pi_{\Lambda_1}(V)$. Adding to $\Lambda_1$ any definition that specializes one of its roots leaves the projection unchanged.
\item $\pi_{\Lambda_1}(V) \cap \pi_{\Lambda_2}(V)$ equals the projection by the type set $D_1 \cap D_2$ exactly when every usage $x \in V$ with $T(x) \cap D_1 \neq \varnothing$ and $T(x) \cap D_2 \neq \varnothing$ also has $T(x) \cap D_1 \cap D_2 \neq \varnothing$.
\end{enumerate}
\end{proposition}

\begin{proof}
For the first claim, $\mathord{\downarrow}(\Lambda_1 \cup \Lambda_2) = D_1 \cup D_2$, and $T(x)$ meets this union exactly when it meets one of the two sets. A port or parameter selected by the union is selected by one of the two root sets, whose incidence closure selects its owner, and the same holds for the ends of a connection, so the union's selection is incidence closed. For the second claim, $\mathord{\downarrow} D_1 = D_1$ because $\preceq$ is reflexive and transitive. For the third claim, the intersection of the two projections is the set of usages of $V$ whose types meet $D_1$ and meet $D_2$. The projection by $D_1 \cap D_2$ is the set of usages of $V$ whose types meet $D_1 \cap D_2$, always a subset of the former. Equality is the stated condition at every usage of $V$.
\end{proof}

The second claim is the statement behind \cref{rsl:sec:packaging} of the article and \cref{rsls:sec:packaging-detail}. A downstream package declares definitions that specialize those of \texttt{ros2-sysmlv2}, and the projection with $R$ unchanged selects its usages as it selects usages of the library's own definitions. The first claim covers a model that uses \texttt{ros2-sysmlv2} beside a second domain library. Each consumer selects by its own roots, and the two selections combine by union. The third claim marks a limit. Take one usage with two types $a$ and $b$ that are incomparable under $\preceq$ and have no common lower type. The root set $\{a\}$ selects it and so does $\{b\}$, so the usage lies in the intersection of the two projections, while the intersected type set $D_1 \cap D_2$ is empty and selects nothing. A consumer that needs the usages that two libraries both read therefore keeps the two predicates and does not intersect the type sets.

The intersection of two incidence closed selections is itself incidence closed, so $\pi_{\Lambda_1}(V) \cap \pi_{\Lambda_2}(V)$ is the set of usages of $V$ that satisfy the predicate $\sigma_{\Lambda_1} \land \sigma_{\Lambda_2}$, an admitted selection in its own right. Singleton typing, $|T(x)| = 1$ for every usage, is sufficient for the condition of the third claim, and unrestricted multiple typing is not, as the example shows.

\section{The Reference Extractor and the Bridge Extractor}
\label{rsls:sec:extractor}

The projection is defined on the evaluated usage set, but a model is written as a tree. The body of the system definition declares usages, and a part usage whose definition declares further usages holds them at nested feature paths, as the Nav2 stack of the hybrid robot model in \cref{rsls:fig:archetype-to-robot}(b) holds its server nodes. An extractor reads this tree. We specify the reading as a two-pass algorithm and show that it computes the projection, and we state how the bridge extractor of the article is checked against it.

Let $\mathcal{H}$ be the containment tree of the model. Its root body is the system definition with its effective members, and each composite part usage in a body contributes a child body, the usage's effective members, at the path of that usage. The effective members of a body are the features of its own declarations and of the definitions it specializes, each once after redefinition. Flattening, $\mathrm{flat}(\mathcal{H})$, gives every usage in every body its full feature path from the root, keeps its types and values, and resolves the owner of every port and parameter and the two ends of every connection to full paths. An end written as a chain of feature names through enclosing parts resolves to the path that the chain spells from the body in which the connection is declared. A boundary delegation, an end that a composite part forwards to a port inside it, must already be lowered to the inner port's path in this representation, and a tree that leaves one unlowered is outside the theorem's premise.

The reference extractor $E_\Lambda$ makes two passes. The first pass indexes every usage of every body of $\mathcal{H}$ by its full path, selected or not, with its types, its values and the resolved paths of its owner or of its ends. The second pass emits the usages with $\sigma_\Lambda(x) = 1$, with the owner and end maps restricted to the emitted set. It stops with an error when the owner of a selected port or parameter, or an end of a selected connection, is not among the emitted usages.

\begin{theorem}[extraction preservation]
\label{rsls:thm:extraction}
Suppose every usage has a unique full path, every connection end resolves to exactly one usage, every boundary delegation that the tree represents has been lowered to this representation, and $\sigma_\Lambda$ is incidence closed on $\mathrm{flat}(\mathcal{H})$. Then
\begin{equation}
E_\Lambda(\mathcal{H}) = \pi_\Lambda(\mathrm{flat}(\mathcal{H})),
\label{rsls:eq:extraction}
\end{equation}
with equal owner and end maps, types and values on every emitted usage. If $\sigma_\Lambda$ is not incidence closed on $\mathrm{flat}(\mathcal{H})$, then $E_\Lambda$ stops.
\end{theorem}

\begin{proof}
We induct on the height of the containment tree. For a body with no child bodies, the first pass indexes exactly its usages at their paths, and the second pass keeps those that satisfy $\sigma_\Lambda$, which is $\pi_\Lambda$ of the flattened body. For a body with children, prefixing the paths of a child's index by the path of the child's usage is injective, and distinct children have distinct paths. By the induction hypothesis every usage below the body is therefore indexed once, at the path that flattening gives it, with the same types and values. The body's own usages are indexed as in the leaf case. A connection declared in the body resolves each end through the full index to one path, whether the end lies in the body or below it, and that path is the one flattening assigns. Hence the index of the first pass is $\mathrm{flat}(\mathcal{H})$ with its maps. The second pass then forms the set of indexed usages that satisfy $\sigma_\Lambda$, which is $\pi_\Lambda(\mathrm{flat}(\mathcal{H}))$. By the first claim of \cref{rsls:prop:laws} the restricted maps take their values in it, so no error is raised and the two sides of \cref{rsls:eq:extraction} are equal with their maps. If instead some selected port or parameter has an unselected owner, or some selected connection an unselected end, the second pass finds that usage absent from the emitted set and stops.
\end{proof}

Normalization is a separate step after extraction. The names of the emitted usages are resolved to ROS2 graph names by one resolver, and two nodes that resolve to one runtime identity are rejected. A deterministic step applied to equal inputs gives equal outputs or the same rejection, so \cref{rsls:eq:extraction} transfers through it, and two extractors that agree on the emitted usages agree on the normalized records that \cref{rsls:sec:conformance} compares.

This transfer shows that two specified computations agree. It does not show that an implementation uses the specified name-resolution or value functions. Canonical endpoint keys can collide even when feature paths are distinct, for two ports at one owner with one role and one resolved channel. The normalizer must either reject such a collision or apply an explicitly declared identification with consistent coordinates, and the comparison of \cref{rsls:sec:conformance} uses rejection.

The bridge extractor of \cref{rsl:sec:generated-package} of the article is the implementation of this specification for models that keep the owner condition of \cref{rsls:eq:closure}. It reads the system through its effective members, so a part or connection that a system inherits from a base system, and a port or parameter that a node inherits from a base definition, is read once after redefinition. It selects a usage when any of its types specializes a root at any depth, so a user's subtype of a library port, interface or connection definition is selected and a definition of the same name in another package is not. It resolves each connection end from its chain of feature names, prefixed by the path of the body in which the connection is declared. It stops with an error when an end of a selected connection does not resolve to a selected port of the role that the library end names, or to a selected frame, and when two nodes resolve to one runtime identity. On one point it departs from $E_\Lambda$, as the article states. It reads ports and parameters from selected nodes only, so a library-typed port on a part that is not a node is left out rather than stopping the extraction, and the owner condition holds for its output by construction. A connection that reaches such a port stops it, as in $E_\Lambda$. For a model that keeps the owner condition the two agree. The regression checks of the bridge cover these cases on two fixture models and their negative controls. They check that a user subtype is selected and a same-named definition elsewhere is not, and that values bound at a usage win over the definition's. They check that inherited parts, connections and ports are read once after redefinition, and that a transform end written as a direct reference resolves to a frame. They check that a duplicate runtime identity and a connection end on a part that is not a node stop extraction with no output written. On the library release that the article describes, the 55 checks pass; they check the bridge extractor against this specification on the two fixture models.

\section{The Bridge Pipeline in Detail}
\label{rsls:sec:pipeline}

\Cref{rsl:fig:pipeline} and \cref{rsl:tab:stages} of the article give the pipeline in outline. This section follows one run of it from the model files to the launched package. It states what the extractor writes, what the generator checks and emits, and how names, parameter values and QoS profiles reach the code. It then describes the lifecycle base class, the activation script and the launch file. The pipeline runs on the modeling host with the Syside Automator SDK and the template engine Jinja2; a ROS2 installation is needed only to build and run the package it writes.

\subsection{Extraction}
\label{rsls:sec:extraction-detail}

The extractor loads the model files together with the library sources through the Syside Automator API, with warnings treated as errors, and finds the system definition named on its command line. It walks the system's containment tree as \cref{rsls:sec:extractor} specifies. It reads the effective members of each body, so that a part, port, connection or parameter inherited from a base definition is read once after redefinition. It resolves every connection end from its chain of feature names, prefixed by the path of the body in which the connection is declared. It classifies each usage by the library definitions that its types specialize, at any depth. It reads values in the scope of the usage, so a topic name, a QoS preset, a parameter default or a frame identifier redefined at a usage wins over the value of its definition.

Its output is one file, \texttt{architecture.json}. For each node it records the name, the namespace, the definition, the archetype the definition specializes, and whether the node is lifecycle-managed. It records the node's publishers and subscribers with their port names, topic names, message types and QoS preset or profile, its service and action servers and clients with their names and types, and its parameters with their names, types and default values. It records each topic connection, service binding and action binding with its two ends as feature paths and the type the two ends share. It records each coordinate frame with its identifier and each transform with its parent, its child and whether it is static. It also records the requirement definitions and usages of the model, from which the generator writes a report beside the package.

The extractor stops, and writes nothing, in four cases. An end of a selected connection does not resolve to a selected port of the role the library end names or to a selected frame. The two ends of a binding carry different message, service or action types. Two nodes resolve to one runtime identity. A node part carries no name. It warns, and continues, when the two ends of a topic connection resolve to different topic names after namespace resolution, and when a bound value cannot be evaluated.

\subsection{Generation}
\label{rsls:sec:generation-detail}

The generator reads \texttt{architecture.json} and renders a package in Python (\texttt{ament\_python} and \texttt{rclpy}) or in C++ (\texttt{ament\_cmake} and \texttt{rclcpp\_lifecycle}) from one set of templates, chosen by a command-line option. Before it writes the first file it checks every record. Each of the following stops generation with an error that names the element. A port without a message type. A parameter with neither a known type nor a default value, or with a default that does not fit the declared type. A QoS profile with a missing field or a depth that is not a nonnegative integer. A node, namespace, topic, service or action name that breaks the ROS2 name rules. Two definitions whose generated file, executable or launch names would coincide. Two generated nodes of one definition. Nothing is written for a model that fails a check.

The package holds, for each node, a base file and a derived file. In Python they are the files \texttt{<node>\_base.py} and \texttt{<node>.py} of the package module. In C++ they are the header \texttt{<node>\_base.hpp}, the source \texttt{<node>\_base.cpp} and the derived source \texttt{<node>.cpp}. The base file is rewritten on every run. The derived file is written only when it does not exist, so the algorithms a developer writes in it survive every later run, and a command-line option discards the derived files when the developer wants them regenerated. The package also holds its manifest and build files, the launch file, the parameter file, the activation script and the conformance monitor, and a report of the model's requirements. When the model declares message types of its own in a package marked as a message package, the generator writes a separate interface package with one \texttt{.msg} file per declared type, named by the model's package name.

\subsection{Names, Parameter Values, and QoS Profiles}
\label{rsls:sec:names-detail}

Names are resolved by one resolver that follows the name rules of \texttt{rcl} for relative, private, and substituted names, so the extractor, the launch file, the parameter file, and the monitor work with the same graph names. A node namespace is made absolute as the node initialization of \texttt{rcl} makes it, a node name must pass the validation of \texttt{rmw}, and the fully qualified name of a node joins the two. A topic, service or action name is expanded as the client libraries expand it when they create an endpoint. A leading tilde is replaced by the node's fully qualified name, the placeholders for the node name and the namespace are substituted, and a relative name is placed under the node's namespace. The expanded name must pass the validation of a full topic name. The status topic of an action is its resolved name followed by the action status suffix, and the monitor finds an action server by that topic. A name that breaks a rule stops generation with the name of the element it belongs to.

The parameter file keys the values of each node by its fully qualified name, the form that the client libraries match. Each default value is rendered in the YAML form of its declared type, so that the parameter parser of \texttt{rcl} reads a string as a string and a number as a number. The generated base class declares each parameter with the type the model gives it.

An endpoint's QoS becomes the client library's profile for the named preset. When the model writes out a profile of its own, it becomes a profile object, in Python and in C++, built from the model's history and depth, reliability, durability, liveliness and the flag that bypasses the ROS2 topic naming conventions. A preset without concrete policies, the system default and the best available, is rendered as the client library's profile of that name, and its policies are then set by the middleware when the endpoint is created. Before writing, the generator compares the two ends of every topic connection on their reliability and durability as the middleware does. A best-effort publisher with a reliable subscriber and a volatile publisher with a transient-local subscriber are reported as incompatible. An end whose preset leaves a policy to the middleware is reported as undetermined until the monitor observes it.

\subsection{The Lifecycle Base Class and the Activation Script}
\label{rsls:sec:lifecycle-base-detail}

For each node the generator emits a lifecycle base class. The base declares the modeled parameters when the node is created, creates the modeled publishers and subscribers in its configure callback, and destroys them on cleanup and on shutdown, so that afterwards the node holds only what it held when it was created. When a callback ends in an error, which leads the node into error processing, the base releases what it created and returns the node to the unconfigured state, from which it can be configured again. Every generated node is lifecycle-managed, as all eight archetypes are. The derived class overrides the hooks that the base calls from its callbacks and the subscription callbacks the base declares, which is where a developer's algorithm lives.

A generated script brings each node to the active state from the state it is in. For each node it reads the current state through the lifecycle services, and requests, one at a time, the transitions that the default state machine of \texttt{rcl} offers from that state on the way to active: configure from unconfigured, activate from inactive, nothing from active. It reads the state again before every request, waits while a node is in a transition state, and requests each transition by its label. It exits with status zero when every node is active at the end, from any starting state, and with status one otherwise, naming each node that is not active.

\subsection{The Launch File and the Wired Mode}
\label{rsls:sec:launch-detail}

The launch file starts every generated node under its namespace and name with the parameter file, a lifecycle manager that configures and activates the generated nodes at start, and the conformance monitor. The manager is the one of Nav2, run with its bond timeout set to zero because the lifecycle nodes of \texttt{rclpy} do not implement the Nav2 bond. A hybrid model whose nodes include Nav2's own servers runs the Nav2 bringup launch and its own manager, with the Nav2 bond timeout, beside this file. The monitor is launched last and reports from its first cycle, five seconds after it starts.

The wired mode is selected on the command line of the pipeline driver. In it, the base class of every node with publishers creates, when the node is activated, one timer per publisher that publishes a default-constructed message of the modeled type, at 10~Hz for an endpoint with the sensor-data preset and at 1~Hz otherwise. It destroys the timers when the node is deactivated, cleaned up, shut down or recovers from an error. Every subscription callback logs the receipt of a message. The whole topology of the model is thus exercisable, and visible to the graph tools of ROS2, before any algorithm exists. The case study of \cref{rsl:sec:case-study} of the article ran the generated package in this mode.

\section{The Generated Monitor: Keys, Groups, and Counting Rules}
\label{rsls:sec:monitor}

The generator renders the monitor from the same records as the package, with the expectations embedded as tables, so a monitor belongs to the model it was generated from. Every five seconds it queries the ROS2 graph through the introspection calls of the client library, which return the node names and namespaces, the topic names and types, the publishers and subscriptions of a topic with their owners and QoS profiles, and the service names. From these it fills one diagnostic status per group, publishes the statuses as one diagnostic array on the topic \texttt{/conformance\_report}, and writes one summary line to its log. It subscribes to no topic of the system and sends no request to any node.

\subsection{The Expectation Tables and Their Keys}
\label{rsls:sec:monitor-keys}

\Cref{rsls:tab:monitor-keys} lists the tables and the key of each entry, which are the keys of \cref{rsls:sec:conformance}. A node is keyed by its name and namespace, which together are its fully qualified name. An endpoint is keyed by its owner, its role and its resolved topic, with the message type as its coordinate. A connection is keyed by its name and carries the topic, the two owners, the message type and the two presets. A QoS expectation is keyed by its topic and carries the reliability and durability of the declared preset. A service and an action are keyed by their resolved names, a parameter by its owner and its name, and a frame by its identifier. The endpoint table carries one entry per modeled endpoint key: every publisher port, every subscriber at a modeled connection on the connection's topic, and every subscriber that no modeled connection reaches, on the topic its own port resolves to and marked as unconnected. An entry whose topic or message type the model leaves empty carries no concrete value.

\begin{table}[!t]
\centering
\footnotesize
\setlength{\tabcolsep}{4pt}
\caption{The expectation tables of the generated monitor, the key of each entry, the coordinates compared at the key, and the graph query that observes it.}
\label{rsls:tab:monitor-keys}
\begin{tabular}{@{}>{\raggedright\arraybackslash}p{2.1cm}>{\raggedright\arraybackslash}p{3.6cm}>{\raggedright\arraybackslash}p{3.4cm}>{\raggedright\arraybackslash}p{4.0cm}@{}}
\toprule
\textbf{Table} & \textbf{Key of an entry} & \textbf{Coordinates} & \textbf{Observed through} \\
\midrule
Nodes & name, namespace & none & the node list \\
Endpoint types & owner, role, resolved topic & message type & the owner's own publisher or subscription record on the topic \\
Connections & connection name & topic, publisher owner, subscriber owner, message type, the two presets & the two owners' records on the topic, with their QoS profiles \\
QoS expectations & topic & reliability, durability of the declared preset & every publisher's profile on the topic \\
Services & resolved service name & none & the service list \\
Actions & resolved action name & none & the action's status topic in the topic list \\
Parameters & owner, parameter name & none & the owner's presence in the node list \\
Frames & frame identifier & none & the transform topics in the topic list \\
\bottomrule
\end{tabular}
\end{table}

\subsection{The Diagnostic Groups}
\label{rsls:sec:monitor-groups}

\Cref{rsls:tab:monitor-groups} lists the ten groups, what each compares, when it reports agreement, and what it reports otherwise. Three rules hold across the groups. A node that runs in the graph but is not in the model is reported at warning level in the node group. A pattern given to the monitor as a parameter excludes the nodes whose full names match it, which is how a simulator bridge or a visualization tool is kept out of the report. A missing node, a wrong message type and a broken connection are reported at error level, and an incompatible pair of observed profiles at an edge is reported at error level as well. A policy that differs from the declared preset, an absent service, action, parameter owner or transform topic, and an out-of-model node are reported at warning level. Each report names the key: the node, the owner and topic of an endpoint, the connection and the end, or the topic, and for a value both the expected and the observed value.

\begin{table}[!t]
\centering
\footnotesize
\setlength{\tabcolsep}{4pt}
\caption{The ten diagnostic groups of the generated monitor. The level is that of the status when the group finds a departure; a group with nothing to report is at the OK level. The last column gives the key and the values that the status names.}
\label{rsls:tab:monitor-groups}
\begin{tabular}{@{}>{\raggedright\arraybackslash}p{2.2cm}>{\raggedright\arraybackslash}p{4.6cm}>{\raggedright\arraybackslash}p{1.3cm}>{\raggedright\arraybackslash}p{5.0cm}@{}}
\toprule
\textbf{Group} & \textbf{What it compares} & \textbf{Level} & \textbf{What a departure names} \\
\midrule
\texttt{node\_presence} & the modeled set of nodes with the node list, in both directions & error; warning for an out-of-model node & each missing node by its full name; each out-of-model node by its full name \\
\texttt{topic\_types} & for each endpoint entry, the message type on the owner's own record & error for a mismatch; warning for an endpoint not yet discovered & role, owner and topic, with the expected and the observed types \\
\texttt{connections} & for each connection, the publisher owner's and the subscriber owner's records on the topic & error & the connection, and which end is not on the topic \\
\texttt{qos\_policies} & for each topic expectation, the reliability and durability of every publisher on the topic against the declared values & warning & the topic, with the expected and the observed policy \\
\texttt{qos\_rxo} & for each connection, the observed publisher profile against the observed subscriber profile, as the middleware matches them & error & the connection, with the two policies \\
\texttt{qos\_divergence} & for each end of each connection, the observed profile against the end's declared preset & warning & the connection and the end, with the declared and the observed policy \\
\texttt{services} & each resolved service name in the service list & warning & the missing service name \\
\texttt{actions} & each action's status topic in the topic list & warning & the missing action name \\
\texttt{parameters} & the owner of each parameter in the node list & warning & the missing owner, with the parameter name \\
\texttt{tf\_frames} & the transform topics in the topic list & warning & the expected frame identifiers \\
\bottomrule
\end{tabular}
\end{table}

\subsection{Counting Rules and the Summary Line}
\label{rsls:sec:monitor-counting}

An expectation with no concrete value is counted as not compared, never as agreeing. The compared count of a group is its number of expectations less those not compared, and the agreeing count is the number of compared expectations that agree. An endpoint that is absent from the graph therefore stays in the compared count and lowers the agreeing count. An endpoint whose declared preset is the best available has no concrete policy to compare and is counted as not compared. The QoS and divergence groups add how many of their not-compared expectations are of this kind, since the middleware negotiates their policies when the endpoint is created. A subscriber that no modeled connection reaches is compared on the type of its own port like a connected one, and the endpoint-type count states how many such subscribers there are and how many of them carry no type. In the divergence group an end without a declared preset is not compared, an end whose declared or observed policy is not concrete is not compared, and an end absent from the graph is counted apart. In the compatibility group a pair with a non-concrete policy on either side is indeterminate and is never reported as incompatible, and an edge with an end absent is unchecked.

The level of the whole report is the highest level among its groups. The summary line that the monitor logs in each cycle gives that level first. It then gives the agreeing and compared counts of the node, endpoint-type, connection, QoS, compatibility and divergence groups, with the not-compared counts in parentheses, in the form
\begin{center}
\texttt{[Conformance WARN] Nodes: 11/11 | Endpoint types: 22/22 (0 not compared) | Connections: 11/11 | QoS: 8/8 (2 not compared) | RxO: 11/11 | QoS-divergence: 18/18 (4 not compared)},
\end{center}
which is the line of the baseline run of \cref{rsls:sec:case-record}. The counts in the summary line are the ones that \cref{rsl:sec:case-study} of the article reports, and the diagnostic array carries the same counts with the keys and values that \cref{rsls:tab:monitor-groups} lists.

\section{Keyed Conformance of the Running Graph}
\label{rsls:sec:conformance}

\Cref{rsl:sec:conformance} of the article compares the model with the running graph on keyed records and states \cref{rsl:prop:conformance}. This section defines the records and their comparison and decomposes the comparison over scopes. It encodes the comparison as a contract on finite facts, so that a monitor that sees the graph through bounds has a sound verdict, and proves the detection bound. It closes with the conditions under which a generated package carries the model's records to the running graph.

\subsection{Records and Their Comparison}
\label{rsls:sec:records}

The ROS2 consumer applies the projection of \cref{rsls:sec:projection} with the roots $R$ and normalizes the selected usages into records, with the resolver of \cref{rsls:sec:names-detail}. A selected node, port, frame or parameter becomes a node, endpoint, frame or parameter record, and a selected connection becomes a relation record keyed by its kind and its ordered end keys. The values read from the usage, the message type of a port and its pinned QoS values among them, become coordinates at that key. The normalizer resolves namespaces and names, retains roles and interface types, and rejects duplicate canonical keys, so two source ports with different paths that meet at one owner, role and channel are reported rather than merged. Write $M$ for the model's record map.

Let $\mathcal{K}$ be a finite catalog of tagged canonical keys. An endpoint key is
\begin{equation}
k(p) = (\mathsf{endpoint}, \operatorname{fqOwner}(p), \operatorname{role}(p), \operatorname{channel}(p)).
\label{rsls:eq:endpoint-key}
\end{equation}
Node and frame keys carry distinct tags. A parameter key contains its owner and parameter name, and a relation key contains its relation kind and ordered end keys. Interface type is a value at an endpoint key, so changing that type is detectable at the same key. A \emph{well-keyed record map} $Y$ has a key set $\mathcal{K}_Y \subseteq \mathcal{K}$ with at most one record at each key. It assigns values $a_Y(k,j)$ to the observable coordinates $j$ of the record at $k$. Raw duplicate keys or unavailable required values prevent admission as a complete record map.

The model record map $M$ pins a coordinate set $J_M(k)$ for each $k \in \mathcal{K}_M$. Define exact record conformance by
\begin{equation}
Y \equiv_M M \quad\Longleftrightarrow\quad \mathcal{K}_Y = \mathcal{K}_M \ \land\ \forall\, k \in \mathcal{K}_M,\ j \in J_M(k),\quad a_Y(k,j) = a_M(k,j).
\label{rsls:eq:record-conformance}
\end{equation}
An unpinned coordinate has no equality obligation. For admitted complete maps, define the discrepancy sets
\begin{align}
\Delta^- &= \mathcal{K}_M \setminus \mathcal{K}_Y, & \Delta^+ &= \mathcal{K}_Y \setminus \mathcal{K}_M,\\
\Delta^a &= \{\, (k,j) : k \in \mathcal{K}_M \cap \mathcal{K}_Y,\ j \in J_M(k),\ a_Y(k,j) \neq a_M(k,j) \,\}.
\label{rsls:eq:record-differences}
\end{align}

\begin{proposition}[exact comparison and unique correspondence]
\label{rsls:prop:comparison}
For complete, well-keyed maps $M, Y$, conformance in \cref{rsls:eq:record-conformance} holds if and only if $\Delta^- = \Delta^+ = \Delta^a = \varnothing$. Equality of a key fixes the only possible correspondence at that key. Each nonempty discrepancy therefore supplies a named inventory or value witness.
\end{proposition}

\begin{proof}
Empty directional differences give both key-set inclusions and hence equality. With this equality, an empty $\Delta^a$ gives every pinned coordinate equality. Conversely, the defining equalities of conformance make all three sets empty. Since each map has at most one record per key, a key-preserving correspondence cannot choose a different image. No inference about the cause of a discrepancy follows from uniqueness of its location.
\end{proof}

The first two sentences of \cref{rsl:prop:conformance} of the article are this proposition read kind by kind, with the model's records as $M$ and the observed records as $Y$. The running graph conforms exactly when the observed and the modeled key sets agree for each kind and every pinned coordinate agrees at its key. A departure is an element of $\Delta^-$, a modeled key with no observed record, of $\Delta^+$, an observed key with no modeled record, or of $\Delta^a$, a pinned coordinate whose two values differ. Each names its key, and an element of $\Delta^a$ names its coordinate as well.

For bounded key operations and a bounded number of compared coordinates per record, constructing \cref{rsls:eq:record-differences} with hash maps takes expected time linear in the number of records. This bound starts with the record inventories. A channel with $p$ publishers and $s$ subscribers can have $ps$ named candidate links, so explicitly constructing its complete relation inventory can already require quadratic output size.

\subsection{Restriction of a Global Comparison}
\label{rsls:sec:restriction}

Let fixed key scopes $\mathcal{K}_1, \ldots, \mathcal{K}_q$ cover the full comparison catalog. Each restricted map inherits the same pinned-coordinate obligations at an overlap key. Every relation key belongs to at least one scope, including dedicated interface scopes when its ends lie in different components. Restrictions are taken from one global model map and one global complete observation. They retain globally computed relation records rather than recomputing links after endpoint restriction.

\begin{proposition}[exact scoped comparison]
\label{rsls:prop:scoped}
Under these coverage and agreement premises,
\begin{equation}
Y \equiv_M M \quad\Longleftrightarrow\quad Y|_{\mathcal{K}_i} \equiv_{M|_{\mathcal{K}_i}} M|_{\mathcal{K}_i} \quad \text{for every } i.
\label{rsls:eq:scoped-comparison}
\end{equation}
\end{proposition}

\begin{proof}
Global key equality gives equality after intersection with each $\mathcal{K}_i$, and global coordinate agreement gives each inherited agreement. Conversely, local equalities and coverage yield
\[
\mathcal{K}_Y = \bigcup_i (\mathcal{K}_Y \cap \mathcal{K}_i) = \bigcup_i (\mathcal{K}_M \cap \mathcal{K}_i) = \mathcal{K}_M.
\]
Every pinned coordinate lies in a covered scope and agrees there, so all global coordinate obligations hold. Covering relation keys is necessary for the same argument to include cross-component relations.
\end{proof}

This proposition decomposes a comparison whose records have already been obtained globally, which is how the generated monitor compares: its groups of \cref{rsls:sec:monitor-groups} are scopes of one observation of the whole graph. Separate publisher and subscriber scopes satisfy key coverage for their endpoints, but each misses the connection between them unless a scope holds both ends, so the relation keys need scopes of their own, as the connection group of the monitor is.

\subsection{Facts and the Contract}
\label{rsls:sec:contract}

The comparison of \cref{rsls:prop:comparison} is two-sided equality on complete maps. A monitor that observes the graph through samples does not hold a complete map at every instant, and its verdict is stated more conveniently on finite facts, with a contract between a required set and a permitted set. The comparison of this subsection fixes a model revision, a deployment scope, a normalization profile, and a finite catalog $V$ of possible named structural facts. The catalog includes primitive facts $V_0$ for node identities, owned endpoints, selected finite attribute values, parameters, frames, and explicit bindings, and derived facts $V_d = V \setminus V_0$ for candidate channel links or participation summaries. Fact atoms encode record presence and finite coordinate alternatives: a primitive record at key $k$ is the presence atom $\operatorname{pres}(k)$, and a value $v$ at its coordinate $j$ is the value atom $\operatorname{val}(k,j,v)$. The shorthand $\operatorname{pub}(a,c,t)$ denotes a publisher record for owner $a$ and channel $c$ with interface-type value $t$. If two value atoms refer to the same key and coordinate, admissibility requires exactly one supported value where that coordinate is mandatory, and a value atom requires the presence atom of its record. Continuous values and unbounded names require a separate finite abstraction before this encoding applies.

A finite positive Horn rule set $H$ over $V$ derives structural relations from declared facts. A rule $A \to r$, with $A \subseteq V$, adds $r$ when all facts in $A$ are present. The principal ROS2 rule schema is instantiated only for catalog endpoints with compatible roles and equal resolved channel and interface type:
\begin{equation}
\{\operatorname{pub}(a,c,t), \operatorname{sub}(b,c,t)\} \longrightarrow \operatorname{link}(a,b,c).
\label{rsls:eq:link-rule}
\end{equation}
Here $\operatorname{link}$ means candidate participation on one channel determined by role, resolved channel, and interface-type equality. The rule does not evaluate QoS compatibility, assert pairwise wiring, establish successful middleware matching, or prove message delivery; a modeled binding stays a primitive fact of its own. Every rule head lies in $V_d$, so no rule creates a node or endpoint declaration. For $S \subseteq V$, let $\operatorname{cl}_H(S)$ be the least set containing $S$ and closed under $H$. Closure is deterministic because the catalog and the rule set are fixed. It is also monotone, so $S \subseteq S'$ implies $\operatorname{cl}_H(S) \subseteq \operatorname{cl}_H(S')$, since every rule whose antecedents hold in $S$ also holds in $S'$ and the finite fixed points are reached by the same rounds.

\begin{definition}[structural contract]
\label{rsls:def:contract}
A finite structural contract is a pair $\mathcal{C} = (L, \Pi)$ of subsets of $V$ with $L \subseteq \Pi$. A closed fact set $T \subseteq V$ satisfies it, written $T \models \mathcal{C}$, when
\begin{equation}
L \subseteq T \subseteq \Pi.
\label{rsls:eq:interval-contract}
\end{equation}
The model supplies the required facts $L$, while $\Pi$ states which additional facts are permitted in the chosen observable scope.
\end{definition}

Both inclusions are needed, because a required endpoint can be absent and a new endpoint can create an unintended candidate link. Exact equality is the special case $L = \Pi$, where both are a closed fact encoding of the model on the chosen comparison coordinates; a permissive requirement uses a larger $\Pi$, and an unrestricted upper bound $\Pi = V$ asks solely for the required facts.

To connect the contract with \cref{rsls:prop:comparison}, let $F(Y) \subseteq V_0$ contain the presence and value atoms of a well-keyed map $Y$, and write
\begin{equation}
\begin{aligned}
L_{\mathrm{rec}} ={}& \{\operatorname{pres}(k) : k \in \mathcal{K}_M\} \cup \{\operatorname{val}(k,j,a_M(k,j)) : k \in \mathcal{K}_M,\ j \in J_M(k)\},\\
\Pi_{\mathrm{rec}} ={}& \{\operatorname{pres}(k) : k \in \mathcal{K}_M\} \cup \{\operatorname{val}(k,j,v) : k \in \mathcal{K}_M,\ j \notin J_M(k) \ \lor\ v = a_M(k,j)\}.
\end{aligned}
\label{rsls:eq:record-interval}
\end{equation}
Both sets are restricted to the declared finite vocabulary, and an unpinned value stays an allowed alternative. For every admitted $Y$,
\begin{equation}
Y \equiv_M M \quad\Longleftrightarrow\quad L_{\mathrm{rec}} \subseteq F(Y) \subseteq \Pi_{\mathrm{rec}}.
\label{rsls:eq:record-encoding}
\end{equation}
The lower inclusion requires every model key and pinned value. The upper inclusion excludes every other key and every different value at a pinned coordinate, and single-valuedness supplies the converse. Since every rule head is a derived fact, closure preserves the primitive part, $\operatorname{cl}_H(F(Y)) \cap V_0 = F(Y)$, so the contract $(L_{\mathrm{rec}}, \Pi_{\mathrm{rec}})$ on the closure of $F(Y)$ decides \cref{rsls:eq:record-conformance}. Separate derived obligations $L_d \subseteq \Pi_d \subseteq V_d$ add relation comparison to it with $L = L_{\mathrm{rec}} \cup L_d$ and $\Pi = \Pi_{\mathrm{rec}} \cup \Pi_d$, and an explicit primitive binding and a derived candidate link keep their different meanings.

\subsection{The Verdict of a Sampled Monitor}
\label{rsls:sec:sampled}

A runtime observer receives a partial and time-dependent view of the graph. To use a structural contract against that view, the observer states which facts it has established and which facts its coverage lets it rule out. A missing response to one graph query is not, by itself, proof that an endpoint is absent.

At a declared sample time $t$, let $D_t \subseteq V_0$ be the normalized primitive facts of the actual in-scope deployment under the chosen catalog and adapter semantics, and let $T_t = \operatorname{cl}_H(D_t)$. The declared profile accounts for every actual in-scope name and supported value. An observed out-of-catalog fact is classified by a predeclared closed-scope policy as a forbidden fact, or the adapter reports that the finite profile does not apply, and in either case it blocks a passing verdict. The observer supplies a lower primitive set $d_t^-$ of established facts and an upper primitive set $d_t^+$ of facts not yet excluded. Its evidence contract is
\begin{equation}
d_t^- \subseteq D_t \subseteq d_t^+.
\label{rsls:eq:primitive-evidence}
\end{equation}
The first inclusion needs correct identification of positive observations. The second needs justified inventory coverage, version alignment, and absence handling, and an incomplete sweep retains every unexcluded catalog fact in $d_t^+$. Both bounds are conditional on a scope and time interpretation. The scope fixes the model and deployment versions, the sample phase, the boundary and the policy that excludes infrastructure records from the compared inventories. A snapshot whose drift during collection is not bounded is not presented as a fact set at $t$.

Positive Horn closure is monotone, so \cref{rsls:eq:primitive-evidence} yields
\begin{equation}
\ell_t = \operatorname{cl}_H(d_t^-) \subseteq T_t \subseteq u_t = \operatorname{cl}_H(d_t^+).
\label{rsls:eq:closed-evidence}
\end{equation}
For a fixed contract $\mathcal{C} = (L, \Pi)$, use the following verdict.
\begin{equation}
\mathsf{Verdict}_t =
\begin{cases}
\mathsf{PASS}, & L \subseteq \ell_t \ \text{and}\ u_t \subseteq \Pi,\\
\mathsf{FAIL}, & L \not\subseteq u_t \ \text{or}\ \ell_t \not\subseteq \Pi,\\
\mathsf{UNKNOWN}, & \text{otherwise}.
\end{cases}
\label{rsls:eq:evidence-verdict}
\end{equation}
The cases do not overlap when $L \subseteq \Pi$ and $\ell_t \subseteq u_t$. The formula is evaluated only when the finite profile is admitted for the observed scope. The bounds carry record and coordinate tags, so a false required fact or an established forbidden fact identifies the affected normalized key.

\begin{theorem}[sound evidence verdict]
\label{rsls:thm:evidence}
When the finite profile accounts for all in-scope actual facts and \cref{rsls:eq:primitive-evidence} holds, a $\mathsf{PASS}$ from \cref{rsls:eq:evidence-verdict} implies $T_t \models \mathcal{C}$, and a $\mathsf{FAIL}$ implies $T_t \not\models \mathcal{C}$. An $\mathsf{UNKNOWN}$ asserts neither conclusion.
\end{theorem}

\begin{proof}
Monotonicity of the closure applied twice to \cref{rsls:eq:primitive-evidence} yields $\ell_t \subseteq T_t \subseteq u_t$. For a $\mathsf{PASS}$, $L \subseteq \ell_t \subseteq T_t \subseteq u_t \subseteq \Pi$, so $T_t \models (L, \Pi)$. For a $\mathsf{FAIL}$ caused by $L \not\subseteq u_t$, choose $r \in L \setminus u_t$. Since $T_t \subseteq u_t$, $r \notin T_t$, and the required inclusion fails. For a $\mathsf{FAIL}$ caused by $\ell_t \not\subseteq \Pi$, choose $r \in \ell_t \setminus \Pi$. Since $\ell_t \subseteq T_t$, the allowed inclusion fails. Both failure reasons may hold, and either suffices. No claim about $T_t$ follows from the $\mathsf{UNKNOWN}$ branch.
\end{proof}

For a required fact $r \in L \setminus u_t$, every deployment compatible with the sound upper bound lacks $r$. For a fact $r \in \ell_t \setminus \Pi$, every compatible deployment contains an unpermitted fact. These are concrete failure witnesses. Under the encoding of \cref{rsls:eq:record-interval} they are the departures of \cref{rsls:prop:comparison}. A required presence atom outside $u_t$ is a modeled key with no observed record. An established presence atom outside $\Pi_{\mathrm{rec}}$ is an observed key with no modeled record, and an established value atom outside $\Pi_{\mathrm{rec}}$ is a pinned coordinate whose observed value differs. The next result identifies exactly what additional deployment constraints can resolve when the interval verdict is unknown.

Let $A_t(D)$ be an independently justified constraint on possible actual primitive deployments. It can express unique values, owner incidence, or a known configuration choice, and it does not assume the architectural contract being tested. Define the compatible completion family and its conformant part by
\begin{align}
\mathcal{E}_t &= \{\, D \subseteq V_0 : d_t^- \subseteq D \subseteq d_t^+,\ A_t(D) = 1 \,\},\\
\mathcal{E}_t^{\mathcal{C}} &= \{\, D \in \mathcal{E}_t : \operatorname{cl}_H(D) \models \mathcal{C} \,\}.
\label{rsls:eq:completions}
\end{align}
The evidence premise is now $D_t \in \mathcal{E}_t$. A nonempty completion family is necessary; an empty family identifies inconsistent evidence or assumptions and yields no verdict.

\begin{proposition}[exact verdict relative to completion evidence]
\label{rsls:prop:completion}
Suppose $D_t \in \mathcal{E}_t$. The verdict
\begin{equation}
\mathsf{Verdict}^{*}_t =
\begin{cases}
\mathsf{PASS}, & \mathcal{E}_t^{\mathcal{C}} = \mathcal{E}_t,\\
\mathsf{FAIL}, & \mathcal{E}_t^{\mathcal{C}} = \varnothing,\\
\mathsf{UNKNOWN}, & \text{otherwise}
\end{cases}
\label{rsls:eq:completion-verdict}
\end{equation}
is sound. In the unknown case, no decision based only on $\mathcal{E}_t$ can soundly assert either conformance or violation for every actual deployment consistent with that evidence. Every definite verdict from \cref{rsls:eq:evidence-verdict} agrees with \cref{rsls:eq:completion-verdict}.
\end{proposition}

\begin{proof}
Membership of the actual deployment in $\mathcal{E}_t$ proves soundness of each universal case. If the verdict is unknown, choose $D_1 \in \mathcal{E}_t^{\mathcal{C}}$ and $D_0 \in \mathcal{E}_t \setminus \mathcal{E}_t^{\mathcal{C}}$. Both satisfy exactly the supplied evidence constraints. Asserting conformance would be false for $D_0$, and asserting violation would be false for $D_1$. Finally, every completion lies between the primitive bounds, so the monotonicity argument in \cref{rsls:thm:evidence} applies to every completion. Its pass condition makes all completions conformant, and either failure witness makes all completions nonconformant.
\end{proof}

The difference can arise from alternatives. Suppose the only rules are $\{a\} \to r$ and $\{b\} \to r$, the contract requires $r$, and $d_t^- = \varnothing$, $d_t^+ = \{a, b\}$. The interval lower closure lacks $r$, so \cref{rsls:eq:evidence-verdict} returns unknown. If independent configuration evidence establishes exactly one of $a, b$, then $\mathcal{E}_t = \{\{a\}, \{b\}\}$, and both completions derive $r$. The exact completion verdict is pass. The configuration evidence is obtained, not assumed from the model's desired availability.

The exact contract $L = \Pi$ on a closed comparison encoding recovers the two-sided comparison of \cref{rsls:prop:comparison} when observation is complete, $d_t^- = d_t^+ = D_t$. Before that point, unknown records remain possible in the upper bound. An unexpected endpoint is reported as a definite violation from sound positive evidence, while absence of a required endpoint needs sound negative coverage. An observed message type or pinned QoS value is compared only if the adapter identifies the value at the same endpoint key. Compatibility between offered and requested QoS is a different predicate from equality with a modeled value, and the generated monitor reports the two in separate groups.

\subsection{The Detection Bound}
\label{rsls:sec:detection}

No finite detection delay follows from \cref{rsls:eq:primitive-evidence} alone. A persistent missing endpoint can remain unknown if its upper bound is never narrowed. The temporal result requires a separate completeness assumption.

\begin{assumption}[bounded observation on a stable scope]
\label{rsls:assum:observation}
The monitor is already sampling when a violation starts. The sample sequence is unbounded in time and has gaps at most $h$. For a fixed admitted profile, whenever the whole relevant primitive deployment is constant on $[t_j - \delta, t_j]$, the adapter supplies exact bounds $d_{t_j}^- = d_{t_j}^+ = D_{t_j}$. The corresponding decision is reported within $\eta$ time units after $t_j$.
\end{assumption}

\begin{theorem}[detection under stable complete acquisition]
\label{rsls:thm:detection}
Under \cref{rsls:assum:observation}, suppose a violation starts at $a$ and the relevant deployment remains constant and violating through the first sample at or after $a + \delta$. Then the report contains a failure by
\begin{equation}
t_{\mathrm{report}} - a \leq \delta + h + \eta.
\label{rsls:eq:detection-bound}
\end{equation}
\end{theorem}

\begin{proof}
Let $t_j$ be the first sample at or after $a + \delta$. The sampling premise gives $t_j \leq a + \delta + h$, and $t_j - \delta \geq a$. The assumed stability therefore covers the entire acquisition window. Exact primitive bounds give $\ell_{t_j} = u_{t_j} = T_{t_j}$. Because that closed fact set violates the contract, it either lacks a required fact or contains a forbidden fact, so \cref{rsls:eq:evidence-verdict} returns failure. Reporting adds at most $\eta$, proving \cref{rsls:eq:detection-bound}.
\end{proof}

The last two sentences of \cref{rsl:prop:conformance} of the article are \cref{rsls:assum:observation} and \cref{rsls:thm:detection} with the exact contract of \cref{rsls:eq:record-interval}. A departure of \cref{rsls:prop:comparison} that starts at $a$ and persists through the first sample taken $\delta$ or more after $a$ is a violation of $(L_{\mathrm{rec}}, \Pi_{\mathrm{rec}})$ at that sample. The failure that \cref{rsls:eq:evidence-verdict} returns names its key by the tag of the witnessing atom. The constants $h$, $\delta$ and $\eta$ are obligations of the monitor and its graph interface. For the generated monitor of \cref{rsls:sec:monitor}, $h$ is its cycle of five seconds plus the duration of one cycle's queries, and $\delta$ is the time the graph discovery of the middleware takes to reflect a change in the records that the queries return.

\subsection{From the Model through the Generated Package to the Running Graph}
\label{rsls:sec:realization}

Let $A = \gamma(M)$ be an emitted software artifact and $D = \operatorname{Recover}(A)$ its normalized declarations. Recovery reads the emitted constructors, launch files, and configuration under a stated supported language fragment. It is a separate computation from generating the expected inventory. Let $R_t$ be a complete runtime record map at a stated time, scope, and version. All three maps use the same keys and model-pinned coordinate policy.

\begin{theorem}[conditional structural realization]
\label{rsls:thm:realization}
If $M, D, R_t$ are complete and well keyed, and
\begin{align}
\mathcal{K}_M &\subseteq \mathcal{K}_D \subseteq \mathcal{K}_M, & \mathcal{K}_D &\subseteq \mathcal{K}_{R_t} \subseteq \mathcal{K}_D,\\
a_M(k,j) &= a_D(k,j) = a_{R_t}(k,j), && k \in \mathcal{K}_M,\ j \in J_M(k),
\label{rsls:eq:realization-premises}
\end{align}
then $R_t \equiv_M M$. More generally, for primitive declarations $D_0$, actual primitive facts $D_t$, and any finite structural contract, if
\begin{equation}
\operatorname{cl}_H(D_0) \models (L, \Pi) \qquad\text{and}\qquad D_t = D_0,
\label{rsls:eq:interval-realization}
\end{equation}
then $\operatorname{cl}_H(D_t) \models (L, \Pi)$. The actual-fact premise can instead be discharged by sound observation bounds whose verdict is $\mathsf{PASS}$ under \cref{rsls:thm:evidence}.
\end{theorem}

\begin{proof}
The first two pairs of inclusions give $\mathcal{K}_M = \mathcal{K}_D = \mathcal{K}_{R_t}$. The coordinate equalities in \cref{rsls:eq:realization-premises} then give \cref{rsls:eq:record-conformance}. For \cref{rsls:eq:interval-realization}, equal primitive inputs give equal least closures under the fixed rule set, so the established contract inclusions transfer to the actual closure. Under observation bounds, \cref{rsls:thm:evidence} supplies those inclusions directly without claiming equality with recovered declarations.
\end{proof}

The premises separate two obligations. Recovery establishes what the generated files declare, including the extension points a developer may use, and the observation establishes what runs in the chosen scope, including declarations created outside the generated files. For the pipeline of \cref{rsls:sec:pipeline}, the base files declare the model's nodes, topic endpoints with their types and profiles, and parameters, keyed and named by the same resolver as the model's records. The monitor of \cref{rsls:sec:monitor} supplies the observation of them. An extension of a generated node preserves the conclusion exactly when the inventories and pinned values it yields remain equal to those of $M$. Any departure is a witness of \cref{rsls:prop:comparison} at a current key, which is what the departures of \cref{rsls:sec:case-record} show.

Pinned profile agreement also transfers predicates on those values. If a static endpoint predicate $\psi$ reads only coordinates in $J_M$, then $R_t \equiv_M M$ implies $\psi(R_t) = \psi(M)$ by substitution of equal arguments. The request-offer comparison of reliability and durability, with best effort below reliable and volatile below transient local, is such a predicate when both coordinates are pinned and observed. A runtime publisher can remain compatible with its subscribers while deviating from the model, so the generated monitor reports the comparison of the two observed profiles at an edge and the divergence of each end from its declared preset as separate groups. Concluding middleware admission additionally requires that this predicate be the actual middleware's admission rule.

\section{The Record of the Case Study}
\label{rsls:sec:case-record}

This section is the record behind \cref{rsl:sec:case-study} of the article. It holds the records of the generated ground robot, the raw lines of the monitor in the baseline run and in each of the three departures, and the runs of the package generated from the same model in Python and in C++.

\subsection{The Model's Records and the Generated Package}
\label{rsls:sec:record-model}

The ground-robot model declares eleven nodes, each specializing one of the library's archetypes, in six namespaces. \Cref{rsls:tab:record-nodes} lists them with the endpoints the extractor read. Every node is lifecycle-managed. The model declares ten requirement definitions, the robot's structure and power system, and the allocation of the nodes to the onboard computer; none of these is selected, and they stay in the model. The projection yields eleven topic connections over nine topics (three subscribers read \texttt{/scan}), one service server, three action servers, two parameters, six coordinate frames and five transforms. The frames are \texttt{map}, \texttt{odom}, \texttt{base\_link}, \texttt{lidar\_link}, \texttt{imu\_link} and \texttt{camera\_link}. The transforms from \texttt{map} to \texttt{odom} and from \texttt{odom} to \texttt{base\_link} are dynamic, and the three sensor mounts from \texttt{base\_link} are static. Ten of the eleven publishers declare a QoS preset, three the sensor-data preset, five the default, one the system default (\texttt{/cmd\_vel\_smooth}) and one the best available (\texttt{/map}), and the mission coordinator's publisher declares none.

\begin{table}[!t]
\centering
\footnotesize
\setlength{\tabcolsep}{4pt}
\caption{The eleven nodes of the ground-robot model as the extractor records them. Message types are those of the ROS2 packages \texttt{sensor\_msgs}, \texttt{nav\_msgs} and \texttt{geometry\_msgs}. A subscriber is listed by the topic of its modeled connection.}
\label{rsls:tab:record-nodes}
\begin{tabular}{@{}>{\raggedright\arraybackslash}p{3.0cm}>{\raggedright\arraybackslash}p{1.9cm}>{\raggedright\arraybackslash}p{3.5cm}>{\raggedright\arraybackslash}p{2.0cm}>{\raggedright\arraybackslash}p{2.7cm}@{}}
\toprule
\textbf{Node} & \textbf{Archetype} & \textbf{Publishes (topic, type, preset)} & \textbf{Subscribes} & \textbf{Also declares} \\
\midrule
\texttt{/sensors/lidar} & \texttt{SensorDriver} & \texttt{/scan}, \texttt{LaserScan}, sensor data & & \\
\texttt{/sensors/imu} & \texttt{SensorDriver} & \texttt{/imu/data}, \texttt{Imu}, sensor data & & \\
\texttt{/sensors/camera} & \texttt{SensorDriver} & \texttt{/camera/image}, \texttt{Image}, sensor data & & \\
\texttt{/perception/}\allowbreak\texttt{obstacle\_detector} & \texttt{Perception\-Pipeline} & \texttt{/detections}, \texttt{PointCloud2}, default & \texttt{/scan}, \texttt{/camera/image} & \\
\texttt{/estimation/}\allowbreak\texttt{ekf\_localizer} & \texttt{Estimator} & \texttt{/odom\_filtered}, \texttt{Odometry}, default & \texttt{/imu/data}, \texttt{/scan} & \\
\texttt{/planning/path\_planner} & \texttt{Planner} & \texttt{/plan}, \texttt{Path}, default & \texttt{/detections}, \texttt{/map} & action server \texttt{compute\_path} \\
\texttt{/control/}\allowbreak\texttt{trajectory\_tracker} & \texttt{Controller} & \texttt{/cmd\_vel}, \texttt{Twist}, default & \texttt{/plan}, \texttt{/odom\_filtered} & action server \texttt{follow\_trajectory}; parameters \texttt{max\_linear\_}\allowbreak\texttt{velocity} 2.0, \texttt{max\_angular\_}\allowbreak\texttt{velocity} 1.5 \\
\texttt{/safety/}\allowbreak\texttt{velocity\_smoother} & \texttt{VelocityFilter} & \texttt{/cmd\_vel\_smooth}, \texttt{Twist}, system default & \texttt{/cmd\_vel} & \\
\texttt{/safety/}\allowbreak\texttt{collision\_guard} & \texttt{VelocityFilter} & \texttt{/cmd\_vel\_out}, \texttt{Twist}, default & \texttt{/cmd\_vel\_smooth}, \texttt{/scan} & \\
\texttt{/map\_server} & \texttt{MapProvider} & \texttt{/map}, \texttt{OccupancyGrid}, best available & & service server \texttt{get\_map} \\
\texttt{/mission\_coordinator} & \texttt{Behavior\-Coordinator} & \texttt{/mission\_status}, \texttt{Path}, none & & action server \texttt{execute\_mission} \\
\bottomrule
\end{tabular}
\end{table}

From these records the generated monitor carries eleven node expectations, 22 endpoint-type expectations (the eleven publishers and the eleven subscribers at the connections), eleven connections, ten QoS expectations, one service, three actions, two parameters and six frames. The generated package holds the eleven base and derived node files, the launch file with the lifecycle manager and the monitor, the parameter file with the two values of the trajectory tracker, and the activation script, as \cref{rsls:sec:pipeline} describes. The service and the action servers are not generated in the version described here, and the monitor reports their absence at warning level.

\subsection{The Baseline Run}
\label{rsls:sec:record-baseline}

The Python package was built with \texttt{colcon} on a Linux workstation with ROS2~Jazzy and the Zenoh middleware (\texttt{rmw\_zenoh\_cpp}), on a ROS domain of its own that held no node before the launch, and launched in the wired mode of \cref{rsls:sec:launch-detail} without a simulator. The lifecycle manager of the launch file brought the eleven nodes to the active state. The monitor wrote twelve summary lines during the run, all with the text
\begin{lstlisting}[language={},basicstyle=\ttfamily\footnotesize,frame=none,aboveskip=4pt,belowskip=4pt]
[Conformance WARN] Nodes: 11/11 | Endpoint types: 22/22 (0 not compared) | Connections: 11/11 | QoS: 8/8 (2 not compared) | RxO: 11/11 | QoS-divergence: 18/18 (4 not compared)
\end{lstlisting}
and \cref{rsls:lst:baseline-report} gives the ten groups of the one diagnostic array captured from the report topic, with the level of each. The two QoS expectations not compared are \texttt{/cmd\_vel\_smooth} and \texttt{/map}, whose presets fix no concrete policy, and the four divergence expectations not compared are the two ends of the connections on those topics. The three out-of-model nodes are the lifecycle manager launched by the package and the two command-line nodes that the run's own graph probes started. The absent service, actions and transform topics are the interfaces the package does not generate. After the teardown no process of the package remained and the node list of the domain was empty.

\begin{figure}[!t]
\begin{lstlisting}[language={},basicstyle=\ttfamily\footnotesize,literate={—}{{---}}1]
node_presence   WARN  All 11 expected nodes present; 3 out-of-model node(s)
                      out_of_model: /lifecycle_manager_custom, /_ros2cli_<pid>, /_ros2cli_daemon_72_<id>
topic_types     OK    All 22 compared endpoint types OK, 0 not compared
connections     OK    All 11 connections verified
qos_policies    OK    8 QoS policies OK, 0 without a publisher, 2 not compared
qos_rxo         OK    All 11 edges RxO-compatible
qos_divergence  OK    No QoS divergence (18 endpoints agree, 0 absent, 4 not compared)
services        WARN  1 services missing, 0 present
                      missing: /get_map
actions         WARN  3 actions missing, 0 present
                      missing: /planning/compute_path, /control/follow_trajectory, /execute_mission
parameters      OK    2 parameter declarations OK (node presence verified)
tf_frames       WARN  TF topics not found — 6 frames expected but /tf and /tf_static not active
                      expected_frames: base_link, camera_link, imu_link, lidar_link, map, odom
\end{lstlisting}
\captionof{lstlisting}{The ten groups of the diagnostic array captured in the baseline run, with their levels, messages and named values as the monitor published them. The process and daemon identifiers of the two command-line nodes are abbreviated.}
\label{rsls:lst:baseline-report}
\end{figure}

\subsection{The Three Departures}
\label{rsls:sec:record-departures}

The departures were introduced in the developer-owned file of the lidar driver alone. A replacement for that file reads an environment variable in its configure hook. When the variable is set, it destroys the publisher that the generated base class created and creates one that departs from the model in one respect. The three departures are the topic \texttt{/scan\_raw} in place of \texttt{/scan}, the message type \texttt{std\_msgs/String} in place of \texttt{LaserScan}, and reliable, transient-local delivery in place of the sensor-data preset. The generated base classes, the launch file and the monitor were the committed ones, and a scratch copy of the package differed from the committed package in that one file only. The package was launched four times for 45~s each on another empty domain, once with the variable unset as the control and once per departure, and the monitor wrote eleven summary lines in each run, each run with one distinct text:
\begin{lstlisting}[language={},basicstyle=\ttfamily\footnotesize,frame=none,aboveskip=4pt,belowskip=4pt]
control  [Conformance WARN] Nodes: 11/11 | Endpoint types: 22/22 (0 not compared) | Connections: 11/11 | QoS: 8/8 (2 not compared) | RxO: 11/11 | QoS-divergence: 18/18 (4 not compared)
topic    [Conformance FAIL] Nodes: 11/11 | Endpoint types: 21/22 (0 not compared) | Connections: 8/11 | QoS: 7/8 (2 not compared) | RxO: 8/11 | QoS-divergence: 15/18 (4 not compared)
type     [Conformance FAIL] Nodes: 11/11 | Endpoint types: 21/22 (0 not compared) | Connections: 11/11 | QoS: 8/8 (2 not compared) | RxO: 11/11 | QoS-divergence: 18/18 (4 not compared)
qos      [Conformance WARN] Nodes: 11/11 | Endpoint types: 22/22 (0 not compared) | Connections: 11/11 | QoS: 7/8 (2 not compared) | RxO: 11/11 | QoS-divergence: 15/18 (4 not compared)
\end{lstlisting}
The control run reproduced the baseline line. \Cref{rsls:tab:record-departures} gives, for each departure, the groups whose status differed from the control's in the diagnostic array captured during the run, with the messages and the named values as published. The other groups, and the services, actions and transform warnings of every run, were as in the control. In the topic departure the publisher is absent from \texttt{/scan}, so the three connections that read the topic are broken and the publisher's endpoint-type expectation is not discovered. The QoS expectation of \texttt{/scan} has no publisher, and the three connection ends at the publisher are absent from the compatibility and divergence groups. In the type departure the publisher is on the topic with the wrong type, and only the endpoint-type group reports it. In the QoS departure the publisher's policies differ from the preset. The QoS group names the topic, and the divergence group names the publisher end of each of the three connections. The observed profiles at the three edges remain compatible, since a reliable, transient-local publisher offers more than the best-effort, volatile subscribers request, so the compatibility group stays at the OK level and the overall level stays at warning, as in the control.

\begin{table}[!t]
\centering
\footnotesize
\setlength{\tabcolsep}{4pt}
\caption{The groups that changed in each departure, with the level, the message and the named values of the diagnostic array captured during the run.}
\label{rsls:tab:record-departures}
\begin{tabular}{@{}>{\raggedright\arraybackslash}p{1.1cm}>{\raggedright\arraybackslash}p{2.2cm}>{\raggedright\arraybackslash}p{1.1cm}>{\raggedright\arraybackslash}p{8.6cm}@{}}
\toprule
\textbf{Run} & \textbf{Group} & \textbf{Level} & \textbf{Message and values} \\
\midrule
topic & \texttt{topic\_types} & WARN & 1 endpoints not yet discovered, 21 OK, 0 not compared; missing: \texttt{pub /sensors/lidar on /scan} \\
topic & \texttt{connections} & ERROR & 3 connections broken, 8 OK; \texttt{scanToDetector}, \texttt{scanToEkf}, \texttt{scanToGuard}: \texttt{pub /sensors/lidar not on /scan} \\
topic & \texttt{qos\_policies} & OK & 7 QoS policies OK, 1 without a publisher, 2 not compared \\
topic & \texttt{qos\_rxo} & OK & 0 RxO-incompatible edge(s), 8 OK, 0 indeterminate, 3 unchecked \\
topic & \texttt{qos\_divergence} & OK & No QoS divergence (15 endpoints agree, 3 absent, 4 not compared) \\
\midrule
type & \texttt{topic\_types} & ERROR & 1 type mismatches, 0 missing, 21 OK, 0 not compared; type\_mismatch: \texttt{pub /sensors/lidar on /scan: expected sensor\_msgs.msg.LaserScan, got ['std\_msgs/msg/String']} \\
\midrule
qos & \texttt{qos\_policies} & WARN & 1 QoS mismatches, 7 OK, 0 without a publisher, 2 not compared; \texttt{/scan}: \texttt{reliability: expected BEST\_EFFORT, got RELIABLE; durability: expected VOLATILE, got TRANSIENT\_LOCAL} \\
qos & \texttt{qos\_divergence} & WARN & 3 endpoint(s) diverge from declared QoS, 15 OK, 0 absent, 4 not compared; \texttt{scanToDetector:pub}, \texttt{scanToEkf:pub}, \texttt{scanToGuard:pub}: \texttt{reliability: declared BEST\_EFFORT, observed RELIABLE; durability: declared VOLATILE, observed TRANSIENT\_LOCAL} \\
\bottomrule
\end{tabular}
\end{table}

\subsection{The Same Model in Python and in C++}
\label{rsls:sec:record-languages}

The packages that the bridge generates from the same model in Python and in C++ were built and run on the workstation in the same way, each on an empty domain of its own. The generated activation script was run after the lifecycle manager had activated the nodes. In both languages the script reported that all eleven nodes were active and exited with status zero, the monitor wrote eighteen summary lines with one distinct text, no package process remained after the teardown, and the node list of the domain was empty. With the bridge as it stood when the monitor compared every modeled endpoint on its owner's record, both languages wrote the baseline line of \cref{rsls:sec:record-baseline}. After the monitor started to count the endpoints of the best-available preset apart and to compare subscribers that no connection reaches, both languages wrote
\begin{lstlisting}[language={},basicstyle=\ttfamily\footnotesize,frame=none,aboveskip=4pt,belowskip=4pt]
[Conformance WARN] Nodes: 11/11 | Endpoint types: 22/22 (0 not compared) | Connections: 11/11 | QoS: 8/8 (2 not compared, 1 negotiated at runtime) | RxO: 10/11 | QoS-divergence: 18/18 (4 not compared, 2 negotiated at runtime)
qos_rxo  OK  0 RxO-incompatible edge(s), 10 OK, 1 indeterminate, 0 unchecked
\end{lstlisting}
where the second line is the compatibility group of the diagnostic array. The one negotiated QoS expectation is \texttt{/map} and the two negotiated divergence expectations are its two connection ends. The graph reports both endpoints of \texttt{/map} with the best-available policies, which the compatibility group counts as indeterminate rather than as agreeing, as \cref{rsls:sec:monitor-counting} states. The node, endpoint-type, connection and QoS counts were the same in both languages and in every run.

\end{document}